\documentclass{aa}  
\usepackage{amsmath,amsfonts,amssymb}
\usepackage[dvipsnames]{xcolor}
\usepackage{color}
\usepackage{url}
\usepackage[varg]{txfonts}
\usepackage{graphicx}
\usepackage{enumitem} 
\usepackage{array}
\usepackage{dblfloatfix}
\usepackage{multicol}

\usepackage{natbib,twoopt}
\usepackage[breaklinks=true]{hyperref} 
\bibpunct{(}{)}{;}{a}{}{,} 
\makeatletter
\newcommandtwoopt{\citeads}[3][][]{\href{http://adsabs.harvard.edu/abs/#3}%
{\def\hyper@linkstart##1##2{}%
\let\hyper@linkend\@empty\citealp[#1][#2]{#3}}}
\newcommandtwoopt{\citepads}[3][][]{\href{http://adsabs.harvard.edu/abs/#3}%
{\def\hyper@linkstart##1##2{}%
\let\hyper@linkend\@empty\citep[#1][#2]{#3}}}
\newcommandtwoopt{\citetads}[3][][]{\href{http://adsabs.harvard.edu/abs/#3}%
{\def\hyper@linkstart##1##2{}%
\let\hyper@linkend\@empty\citet[#1][#2]{#3}}}
\newcommandtwoopt{\citeyearads}[3][][]%
{\href{http://adsabs.harvard.edu/abs/#3}
{\def\hyper@linkstart##1##2{}%
\let\hyper@linkend\@empty\citeyear[#1][#2]{#3}}}
\makeatother

\usepackage[breaklinks=true]{hyperref} 
\hypersetup{colorlinks=true,linkcolor=blue,citecolor=blue,urlcolor=blue}

\newcommand*\rot{\rotatebox{90}}

\begin{document}

\title{Multi-season optical modulation phased with the orbit of the super-Earth 55 Cnc e}

   \author{S. Sulis\inst{1}
          \and
          D. Dragomir\inst{2,3}
          \and
          M. Lendl\inst{1,4}
          \and
          V. Bourrier\inst{4}
          \and
          B. O. Demory\inst{5}
          \and
          L. Fossati\inst{1}
          \and
          P. E. Cubillos\inst{1}          
          \and
          D. B. Guenther\inst{6}
          \and
          S. R. Kane\inst{7}
          \and
          R. Kuschnig\inst{8}
          \and
          J. M. Matthews\inst{9}
          \and
          A. F. J. Moffat\inst{10}
          \and
          J. F. Rowe\inst{11}
          \and
          D. Sasselov\inst{12}
          \and
          W. W. Weiss\inst{8}
          \and 
          J. N. Winn\inst{13}
          }

   \institute{Space Research Institute, Austrian Academy of Sciences, 
Schmiedlstra{\ss}e 6, 8042 Graz, Austria\label{inst1} \\
   \email{sophia.sulis@oeaw.ac.at}
         \and
   Department of Physics and Kavli Institute for Astrophysics and Space Research, Massachusetts Institute of Technology, Cambridge, MA 02139, USA
          \and
   Department of Physics and Astronomy, University of New Mexico, Albuquerque, NM, USA
      \and
      Observatoire de l'Université de Genève, 51 chemin des Maillettes, 1290 
Sauverny, Switzerland
        \and
        University of Bern, Center for Space and Habitability, Sidlerstra{\ss}e 
5, CH-3012 Bern, Switzerland
         \and
        Institute for Computational Astrophysics, Department of Astronomy and Physics, Saint Mary’s University, Halifax, NS B3H 3C3, Canada
         \and
         Department of Earth and Planetary Sciences, University of California, Riverside, CA 92521, USA
         \and
        Institut f{\"u}r Astronomie, Universit{\"a}t Wien T{\"u}rkenschanzstrasse 17, A-1180 Wien, Austria
         \and
        Department of Physics and Astronomy, University of British Columbia, 6224 Agricultural Road, Vancouver, BC V6T 1Z1, Canada
                \and
        Observatoire Astronomique du Mont Mégantic, Departement de Physique, Université de Montréal C. P. 6128, Succursale:Centre-Ville, Montréal, QC H3C 3J7, Canada 
         \and
        NASA-Ames Research Park, Mail Stop 244-30, Moffett Field, CA 94035-1000; 
         \and
        Harvard-Smithsonian Center for Astrophysics, 60 Garden Street, Cambridge, MA 102138
          \and
    Department of Astrophysical Sciences, Princeton University, 4 Ivy Lane, Princeton, NJ 08544, USA
    }

   \date{Received XX, 2019; accepted XX, 2019}


  \abstract
   {55 Cnc e is a transiting super-Earth orbiting a solar-like star with an orbital period of $\sim17.7$ hours.
   In 2011, using the Microvariability and Oscillations in Stars (MOST) space telescope, a quasi-sinusoidal modulation in flux was detected with the same period as the planetary orbit. The amplitude of this modulation was too large to be explained as the change in light reflected or emitted by the planet.
   }
   {The MOST telescope continued to observe 55 Cnc e for a few weeks per year over five years (from 2011 to 2015), covering $143$ individual transits. This paper presents the analysis of the observed phase modulation throughout these observations and a search for the secondary eclipse of the planet.
}
   {The most important source of systematic noise in MOST data is due to stray-light reflected from the Earth, which is modulated with both the orbital period of the satellite ($101.4$ minutes) and the Earth’s rotation period.  We present a new technique to deal with this source of noise, which we combined with standard detrending procedures for MOST data. We then performed Markov Chain Monte Carlo analyses of the detrended light curves, modeling the planetary transit and phase modulation.
}
{  
We find phase modulations similar to those seen in 2011 in most of the subsequent years; however, the amplitude and phase of maximum light are seen to vary, from year to year, from $113$ to $28$ ppm and from $0.1$ to $3.8$ rad. The secondary eclipse is not detected, but we constrain the geometric albedo of the planet to less than $0.47$ ($2\sigma$). 
}
   {While we cannot identify a single origin of the observed optical modulation, we propose a few possible scenarios. Those include star-planet interaction, such as coronal rains and spots rotating with the motion of the planet along its orbit, or the presence of a transiting circumstellar torus of dust. However, a detailed interpretation of these observations is limited by their photometric precision. 
   Additional observations at optical wavelengths could measure the variations at higher precision, contribute to uncovering the underlying physical processes, and measure or improve the upper limit on the albedo of the planet.
}


\maketitle

\section{Introduction}

Though super-Earths as a category of exoplanets have been discussed for nearly a decade, the nature and origins of these planets are diverse 
\citepads{Rogers2011,Hansen2012, Mordasini12,  Chiang2013, Batygin2015, Dorn18}. 
A wide range of compositions are possible for these planets, whose mass and size 
lie between those of the Earth and Neptune 
\citepads{2008ApJ...673.1160A,Rogers1}.
Super-Earths are among the most numerous planets within the sample of detected
planets, even though there are no analogs to these planets 
in our solar system \citepads{Mayor11,Petigura13, Fressin13}. 
Recent studies found that, within this size regime, the planet size distribution is bimodal and has a gap between 1.5 and 2.0\,R$_{\oplus}$ \citepads{Fulton17, Fulton_2018}. The most widespread explanation for this ``radius gap'' is photoevaporation \citepads{Lecavelier_2007,Davis_2009,Ehrenreich_2011}. 
In this scenario, planets for which the H/He atmosphere
constitutes less than about 1\% of the total mass at the time of the
dispersal of the protoplanetary disk are fated to lose this atmosphere completely within about 100 Myr. In contrast, planets with initially more massive atmospheres are able to retain enough gas to cause the mean density to be substantially lower than that of a purely rocky planet.
The former have completely lost their primary, hydrogen-dominated atmosphere, and therefore their (small) radius depends exclusively on the average density of the rocky core. For the latter, instead, because of the low planetary mass (i.e., low gravity) and low atmospheric mean molecular weight, the envelope extends far from the rocky surface leading to a larger planetary radius \citepads{owen2017,jin2018,Eyl17}.
The planet 55 Cnc e itself is located at a very short orbital distance and exposed to more intense stellar radiation than the bulk of planets making up the populations discussed above. This planet is an example of only a handful objects with low masses and high irradiation often referred to as ultra-short period planets (USPs).

With an orbital period of just $0.736$ days, the super-Earth 55 Cnc e is an extremely hot, presumably tidally locked super-Earth with a brightness temperature of $\sim 2700~\pm 270$ K \citepads{Demory_2016_nature}. It orbits the third brightest star ($V$ mag = 5.95; {\it TESS} mag = 5.48) known to host a transiting exoplanet in this size category \citepads{2011ApJ...737L..18W, Dem11} after HD 39091 \citepads{Hugh_2002,Gandolfi_2018}
and HD 219134 \citepads{Motalebi_2015, Vogt_2015}.
Based on transit and radial velocity data, the radius of 55 Cnc e is $1.88~\pm~0.03$ R$_\oplus$ and the mass is $8.0~\pm~0.3$ M$_\oplus$ 
\citepads{2018A&A...619A...1B}. The high bulk density is $6.7~\pm~0.4$ g cm$^{-3}$.
Given these measurements, it is not yet possible to tell whether the planet has an Earth-like composition (an iron core surrounded by a silicate mantle) or a rocky core with an envelope of volatiles. 
Atmospheric escape models for such a small, highly irradiated planet predict
that there should be no substantial H/He envelope \citepads{Gillon2, 
Demory_2016_nature, Kubyshkina18} --- and indeed no hydrogen exosphere has been detected \citepads{Ehr12}. However, given its bulk density, the planet is most likely surrounded by a heavyweight atmosphere \citepads{2018A&A...619A...1B}.

This scenario is supported by Spitzer $4.5$ $\mu$m phase curve observations \citepads{Demory_2016_nature}. These observations show an eastward offset of the hot spot of the planet and a night temperature of $1380~\pm~400$ K; both of these characteristics require some heat circulation. 
If 55 Cnc e were a ``lava planet" with no atmosphere, it is unlikely that its heat redistribution efficiency would be sufficiently high to explain these two 
features \citepads{Kite16, Ang17}. It seems more likely that 55 Cnc e
has an optically thick atmosphere, as suggested by \citetads{Demory_2016_nature},  \citetads{Ang17}, and later by \citetads{2018A&A...619A...1B} through refined values of the planet mass and radius.
The presence of a high-metallicity atmosphere is also indicated by the tentative detection of Ca$^+$ and Na in the planet exosphere \citepads{Rid16}. The present-day atmosphere may well be shrouding a molten surface, and might have originated from volcanic outgassing.

A quasi-sinusoidal modulation in the optical flux of the 55 Cnc e system was detected using data from the Microvariability and Oscillations in Stars (MOST) space telescope (\citeads{2011ApJ...737L..18W}, hereafter W11). The modulation had the same period as the planet and an amplitude initially measured at $168~\pm 70$ ppm. While small, this amplitude is too large to be due to only scattered light from the planet, which cannot exceed $\sim30$ ppm in the MOST bandpass. For the more massive giant planet HD 20782b, similar MOST observations revealed the signature of reflected light from the planetary atmosphere as the planet passed through periastron \citepads{2016ApJ...821...65K}.
Therefore, W11 suggested instead that the modulation was some sort of instrumental artifact, or that it may represent a previously unknown type of star-planet interaction. 

Since 55 Cnc e is not sufficiently massive to give rise to observable signatures of tidal interaction in the MOST photometry, the flux modulation at the period of 55 Cnc e may indicate the existence of magnetic interactions between the planet and its host star.
A scenario involving interactions between the stellar corona activity and the planet has already been proposed by \citetads{2018A&A...615A.117B} to explain some modulation in flux observed in UV data as well. Located in a very close orbit, 55 Cnc e is an ideal target for detecting interactions that consist of an active region on the stellar surface rotating  with the planet orbital motion instead of the stellar rotation period \citepads{Shkolnik_2003,Walker_2008,Poppenhaeger_2011,Lanza_2012,Strugarek_2015,Shkolnik_2018, Wright_2015,Cauley_2018,Strugarek_2019}.

Other analyses for 55 Cnc e, based on IR observations of the secondary eclipse have indicated variability in its secondary eclipse depth 
\citepads{Demory_2016,Tam18}. The authors found a $4\sigma$ difference in eclipse depth between two epochs of observations (acquired in 2012 and 2013, respectively). 
\citetads{Demory_2016} suggested either volcanic activity or an inhomogeneous circumstellar torus of gas and dust as possible explanations.
Alternatively, \citetads{Tam18} proposed that the planet may be intermittently covered by reflective grains originating from volcanic activity or cloud variability. 

55 Cnc e has already benefited from numerous multiwavelength observations in the IR \citepads{Dem11, Demory_2016, Demory_2016_nature}, UV \citepads{2018A&A...615A.117B}, and optical (W11; \citeads{Gillon2,2014IAUS..293...52D}). In this paper we present an extensive dataset of optical photometry obtained with MOST, most of which has not been previously published. We use these data to provide new clues to the nature of this mysterious planet, by monitoring the time variable nature of this system, constraining the albedo of planet e and searching for transits of the 
other four known planets in the system. We describe the MOST observations and 
their reduction in Sections~\ref{sec:obs} and \ref{sec:analysis}. The data analysis, results on the refined transit parameters, and discussion regarding 55 Cnc e are found in Sections~
\ref{sec4} and \ref{sec:disc}, respectively. In Section~\ref{sec:other} we 
present a search for transits of the other known planets in the system, and we 
conclude in Section~\ref{sec:conc}.


\begin{table}\centering
\caption{Dates, durations, exposure times, and initial number of transits ($N_{tr}$) corresponding to the MOST observations taken between 2011 and 2015. Note: Dates are given in Barycentric Julian Days [BJD] to within a constant $2~450~000$. }
{\setlength{\extrarowheight}{3pt}
\begin{tabular}{|c|c|c|c|c|c|} 
\hline
Standard  & \multicolumn{2}{|c|}{Dates [BJD]} &   Duration &  Integration  & $N_{tr}$ \\
 \cline{2-3}
Year  &  Start  & End  &[days] &  time [s]  & \\
 \cline{2-3}
\hline \hline
2011 &$5599.52$ & $5614.50$ & $14.98$ & $40$ & 18 \\
2012 &$5940.51$ & $5982.67$ & $42.16$ & $40$  & 52 \\
2013 &$6328.78$ & $6348.70$ & $19.93$ & $40$  & 27 \\
2014 &$6689.10$ & $6713.52$ & $24.43$ & $60$  & 25 \\
2015 &$7033.64$ & $7074.50$ & $40.86$ & $120$ & 53 \\
\hline
\end{tabular}}
\label{tab:obs}
\end{table}


\begin{figure*}[t!]
   \resizebox{\hsize}{!}{\includegraphics{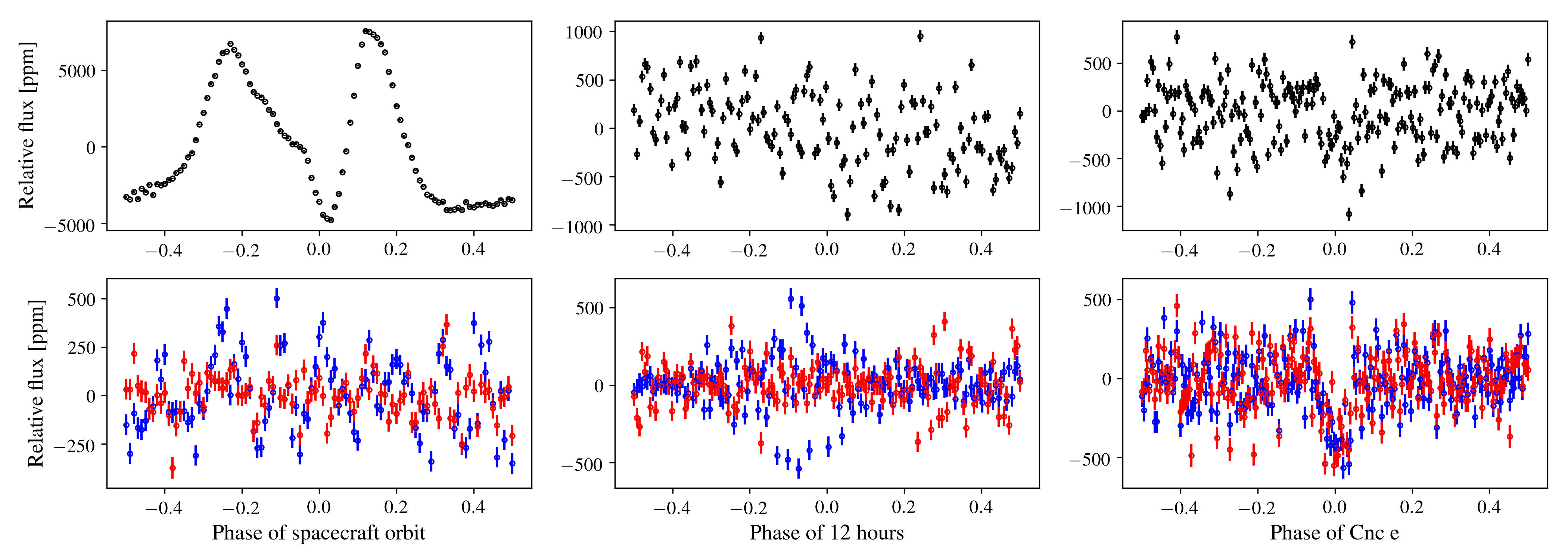}}
    \caption{MOST data obtained in 2015 after pre-whitening (top panels, see Sec.~\ref{sec31}) and after the correction of the Earth stray-light variations (bottom panels, see Sec.~\ref{sec32}). We show the relative flux phase folded at the satellite period (left, binned into $1$-minute), a period of $12$ hours (middle, binned into $5$ min), and the planet period (right, binned into $5$ min). In the bottom panels, observations detrended by the classical procedure (see Sec.~\ref{sec321}) are shown in blue and by the time-shift procedure (see Sec.~\ref{sec322}) in red. In these plots, the uncertainties are based on the original unscaled photometric uncertainties.}
    \label{Fig_steps}
\end{figure*}


\section{MOST observations of 55 Cnc}
\label{sec:obs}

The MOST telescope \citepads{2003PASP..115.1023W,2004AAS...20513401M} is a now inactive microsatellite carrying a $15$ cm optical telescope, which acquires light through a broadband filter spanning the visible wavelengths from $350$ to $700$ nm. This instrument remains in a Sun-synchronous polar orbit with a period of $101.4$ minutes, which allowed it to monitor stars in a continuous viewing zone (CVZ) without interruption for up to eight weeks. The CVZ covered a declination range from $+36^\circ>\delta>-18^\circ$. 
Stars brighter than V$\sim~5-6$ were observed using a Fabry microlens to project 
onto the CCD an image of the telescope pupil illuminated by the target. Fainter 
stars were observed in direct imaging mode, in which the defocused images of the 
stars were projected onto the CCD \citepads{2006ApJ...646.1241R}.

55 Cnc was in the CVZ of MOST, and was observed every year from 2011 to 2015 with 
timespans between about $15$ and $42$ days. The observations were acquired in 
direct imaging mode with an exposure time of $0.5$ s per individual frame. The 
images were downloaded from the satellite in stacks of 40 to 240, resulting in total 
integration times ranging from $20$ to $120$ s (excluding overheads) per downloaded data point.
Including overheads, the sampling rate ranged from $20.91$ to $124.44$ s.
Table \ref{tab:obs} shows the dates, duration, and exposure times used for each of the five time series.

Raw light curves were extracted from the images using aperture photometry. We found that an aperture radius of five pixels almost always gave the lowest scatter in the residuals, so for consistency we used this size to extract all of the MOST photometry. We detail the reduction and analysis of the light curves in the next section.


\section{Photometric analysis}
\label{sec:analysis}

There are several challenges in reducing MOST observations to obtain the final light curves  \citepads{2006ApJ...646.1241R}. In the steps detailed below, we independently reduce each of the five datasets taken 
between 2011 and 2015. We note that the first two datasets (2011-2012) have already been  presented in W11 and \citeads{2014IAUS..293...52D}; hereafter D14). The detrending steps used in the present study follow established methods used to reduce MOST datasets (\citeads{2006ApJ...646.1241R}, W11, D14).

\subsection{Data pre-whitening}
\label{sec31}
The first step consists in removing the extreme outliers exceeding ten standard 
deviations ($\sigma$) from the median flux. 
To obtain a homogeneous time series for each dataset, and also because we observed correlations between the significant outliers and different integration times, we removed data points with integration times differing from the values given in Table \ref{tab:obs}. \\
For each year, we fit the entire dataset ($15$ to $42$ days) with a fifth degree polynomial function of both the sky background and pixel-to-pixel shifts.
We then divided the time series by the best-fit polynomials to obtain a corrected, normalized sequence.
Then, we again eliminated outliers lying above $5\sigma$ from the median flux ($<0.5\%$ of the time series).

After these steps, we removed some parts of the light curves that are affected by large observational gaps (mainly due to tracking lost). We find that doing so increases the signal-to-noise ratio (S/N) of the detected transits. Indeed, observations surrounding these data gaps show a particularly large number of outliers compared to the remaining  values.
For the 2011 dataset, we removed the first $0.6$ and last $2.1$ days (as done in W11), for 2012 we removed the first $0.5$ and last $5.85$ days, for 2013 only the last $2.67$ days, for 2014 the first $8.31$ and last $0.643$ days, and for 2015 the first $10.71$ days (see Appendix.~\ref{App1}).

In the resulting sequences, we observed a long-term variation that may be due to stellar activity; the stellar rotation period is $38.8$ days
(\citeads{2018A&A...619A...1B}). This variation is shown and discussed in Appendix.~\ref{App1}. This had to be corrected before we could make the final correction of patterns related to the satellite motion (see Sec.~\ref{sec32}). We proceeded as follows: we first masked transits and secondary eclipse, and then binned the observations into intervals of twice the planet orbital period. Then, we fit a spline function to the binned data, and removed it from the initial unbinned sequence. We have 
investigated several alternatives to this technique (e.g., boxcar, Gaussian or 
\citeads{savitzky64} filters) but, as long as we consider filter sizes larger 
than twice the planetary period, all of these approaches lead to similar results.

\subsection{Correction for stray-light flux variations}
 \label{sec32}

The MOST satellite completed one polar orbit around Earth in $P_{sat} = 
101.4$ min.
Its observations are affected by the scattered Earthshine, which 
generates flux variations with amplitudes that vary from orbit to orbit 
depending on the part of the Earth visible to the satellite.
These variations are modulated with both the orbital period of the satellite and 
the Earth's rotation period. 
An example of a pattern observed at $P_{sat}$ is shown in the top left panel 
of Fig.~\ref{Fig_steps}, while the sinusoidal-like pattern observed at the 
Earth's rotation period is shown in the top middle panel. 
As discussed in W11, the stray-light timescales and amplitudes are too different from the orbital period of planet e to mimic the variations observed at the planetary period (see Sec.~\ref{sec42}).
However, the orbital period of the satellite is comparable to the planet transit duration ($\approx 95$ min) and the correction of the stray-light patterns should be done carefully to avoid any influence on the inferred transit parameters.  

\subsubsection{Classical method}
\label{sec321}

\begin{figure}[!t]\centering
\resizebox{0.92\hsize}{!}{\includegraphics{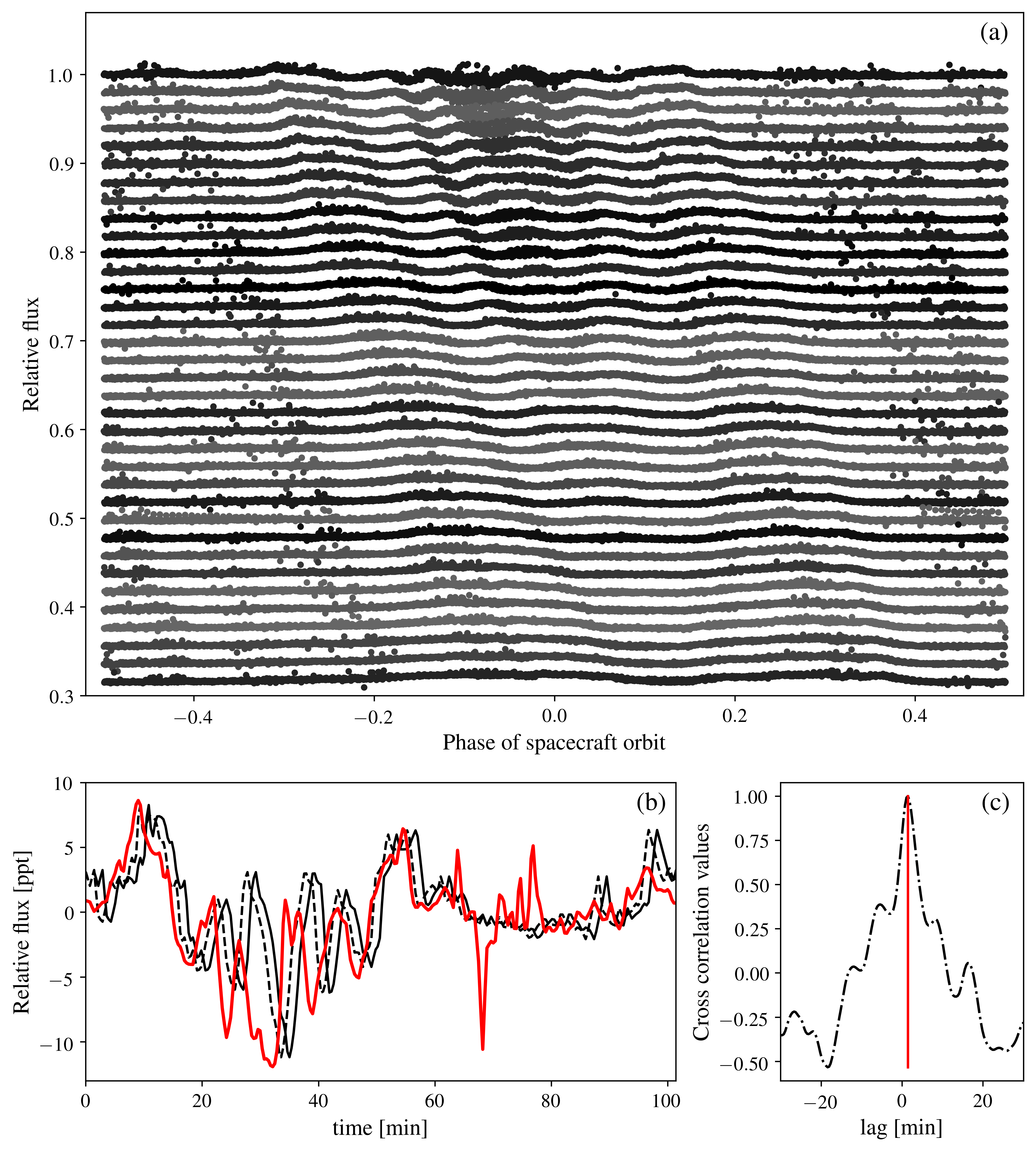}}
    \caption{Illustration of the Earth stray-light variation with respect to the orbital period of the satellite (2012 dataset). (a) Relative flux of the one-day sequences phase folded at the MOST satellite period and binned over $5$-min intervals. The different series have been shifted in flux for visibility. The shape of the stray-light evolves from day-to-day and a time delay is observed. 
(b) Example of patterns with a duration of one satellite period belonging to the same one-day time series. The y-axis is in part-per-thousand. The solid lines show an unshifted (black) and the reference (red) pattern. The dashed line shows the black pattern corrected by the estimated time delay (-$86$s), which was found using the cross-correlation function shown in (c).}
    \label{Fig_tshift}
\end{figure}

The shape of the variations induced by stray-light is variable from one satellite orbit to another. In panel (a) of Fig.~\ref{Fig_tshift}, we show the one-day sequences of the 2012 dataset phase folded on the satellite orbital period.
Each of these shorter sequences contains approximately $14 ~ P_{sat}$. We 
observe a variability in the shape of the pattern and a time delay between the distinct features. These variations illustrate a need to treat MOST observations on a similar timescale.
Traditional techniques (\citeads{2004PASP..116.1093R, Row08, Dra13}) consist in correcting these stray-light systematics on short sequences (e.g., of two-day length). Typically, the short time series are phase folded at the satellite orbital period and a moving average filter is iteratively removed for each of them. 
In this paper, we performed a similar technique to correct for the stray-light variations, which is detailed below:
\begin{enumerate}
\item Select a one-day time-series (corresponding to $\sim 14~P_{sat}$)
\item Mask transits and eclipses
\item Phase fold on satellite orbital period
\item Use a Savitzky-Golay (SG) filter of window size $w_1$ (instead of a 
moving average) to model the detailed variability at the MOST period
\item Remove this variability from the entire one-day sequence (including transits and eclipses)
\item Unfold the subseries
\item Repeat steps 1 to 6 for each one-day sequence
\end{enumerate}

Then, as some correlated noise remains, we removed the final structures related to the orbital period of the satellite and the Earth's rotation period as follows:
\begin{enumerate}\addtocounter{enumi}{7}
\item Mask transits and occultations from the entire light curve
\item Phase fold the series at the MOST orbital period
\item Use a SG filter (of width $w_2$) to model the residual variability at the 
satellite period
\item Remove this variability from the entire light curve (including transits 
and eclipses)
\item Repeat steps 8 to 11 for the satellite period harmonic $P_{sat}/2$ 
(using width $w_2$), the Earth's rotation period and its $12$-hour harmonic 
(using width $w_3$)
\item Unfold the series and remove the final $3\sigma$ outliers from the median 
value
\end{enumerate}

We chose the different parameters involved in this procedure (length of the subseries, 
window sizes of the filters, and sigma clipping level) such that they 
maximize the final S/N of the known planetary transits (evaluated using Equation (2) of 
\citeads{2006MNRAS.373..231P}).
Table~\ref{table0} lists the best window widths $\{w_1, w_2, w_3\}$ found during 
this procedure, the final root-mean-square (rms) measured out-of-transit and the transit S/N for the different sequences (first horizontal box). The final length of the observations ($
T_{obs}$), the number of data points ($N_{pts,}$), and the number of transit 
events ($N_{tr}$) are also indicated at the end of the table. When the 
datasets are combined, we count $157~770$ data points and $143$ transit events. 

The bottom panels of Fig.~\ref{Fig_steps} show this light curve detrending procedure (blue lines), illustrating how stray-light variations related to $
P_{sat}$ (left) and the Earth rotation (middle) are removed.
 Fig.~\ref{Fig_lcs_all} shows the combined light curve phase folded at the planetary orbital period and Fig.~\ref{Fig_lcs} shows each sequence between 2011 and 2015 (left column).

The detrended light curve corresponding to the 2011 dataset (top left panel of Fig.~\ref{Fig_lcs}) is comparable with that reduced by W11 using a similar detrending technique based on moving average filters. However, we obtain a slightly higher scatter than W11 at the transit location owing to the transit masking steps added to the procedure described above (steps 3 and 8).

\begin{table}[!t]\centering
\caption{Parameters involved in the classical and time-shift detrending procedures. Note: $\dagger$ SG window widths have to be uneven, the rms units are ppm, and $ T_{obs}$ is in days.}
{\setlength{\extrarowheight}{1pt}
\begin{tabular}{|c|c|c|c|c|c|c|}
\cline{2-7} 
\multicolumn{1}{c|}{} &Year & 2011 & 2012 & 2013 &2014 &2015 \\
\cline{2-7} \hline
&$w_1~^\dagger$  & $27$ & $7$ & $7$ & $27$ & $7$ \\
&$w_2~^\dagger$  & $331$ & $681$ & $837$ & $341$ & $801$ \\
&$w_3~^\dagger$  & $357$ & $751$ & $957$ & $217$ & $57$ \\
\rot{\rlap{~Classic}}
&$\rm rms$ & $1016$ & $1031$ & $722$ & $915$ & $1213$ \\
&$\rm S/N$  & $13.9$ & $26.7$ & $20.9$ & $15.9$ & $10.2$ \\
\hline  \hline 
&$w_1~^\dagger$  & $27$ & $39$ & $91$ & $43$ & $27$ \\
&$w_2~^\dagger$  & $261$ & $861$ & $131$ & $321$ & $711$ \\
&$w_3~^\dagger$  & $2111$ & $121$ & $2031$ & $671$ & $171$ \\
&$\rm rms$       & $995$ & $1084$ & $773$ & $561$ & $1162$ \\
\rot{\rlap{~Time-shift}}
&$\rm S/N$  & $13.9$ & $26.8$ & $21.2$ & $16.4$ & $9.4$ \\
\cline{1-7} \cline{2-7}
\multicolumn{1}{c|}{} &$N_{pts}$  & $21~236$ & $67~445$ & $32~521$ & 
$17~811$ & $18~757$ \\
\multicolumn{1}{c|}{} &$T_{obs}$ & $12$ & $35$ & $17$ & $14$ & $29$ 
\\
\multicolumn{1}{c|}{} &$N_{tr}$  & $16$ & $47$ & $23$ & $18$ & $39$ \\
\cline{2-7} 
\end{tabular}}
\label{table0}
\vspace{-0.25cm}
\end{table}


\subsubsection{Improvement of the traditional detrending method: Time-shift procedure}
\label{sec322}

We propose an alternative method to improve the modeling of the stray-light pattern modulated at the satellite orbital period (step 4 of the classical procedure). 
As shown in panel (a) of Fig.~\ref{Fig_tshift}, we observe a variability in both the shape of the pattern and a time delay between the distinct features. 

However, smaller time delays are also present between individual orbits during the one-day sequences shown in Fig.~\ref{Fig_tshift}. To take these into account, we developed a new technique that consists in 
cross-correlating each of the $14$ individual $P_{sat}$ features with a 
reference sequence; because of small data gaps we used the sequence with the 
largest number of data points. During this step, both transits and eclipses are 
masked. After compensating the time delay for each of the sequences, we isolated 
the general pattern using a SG filter (width $w_1$) and removed it from the considered one-day time series (containing transits and eclipses). Then, we applied 
steps 5) to 12) of the classical procedure described above. 
An example of the time delay between two consecutive patterns is shown by the 
black and red solid lines in panel (b) of Fig.~\ref{Fig_tshift}. A 
normalized cross-correlation function derived from two of these short series is shown in panel (c), where we find a time shift of $-86$ seconds. The dashed black curve in panel (b) shows the black feature shifted by this delay to match the reference feature.  

\begin{figure}[!t]\centering
\resizebox{\hsize}{!}{\includegraphics{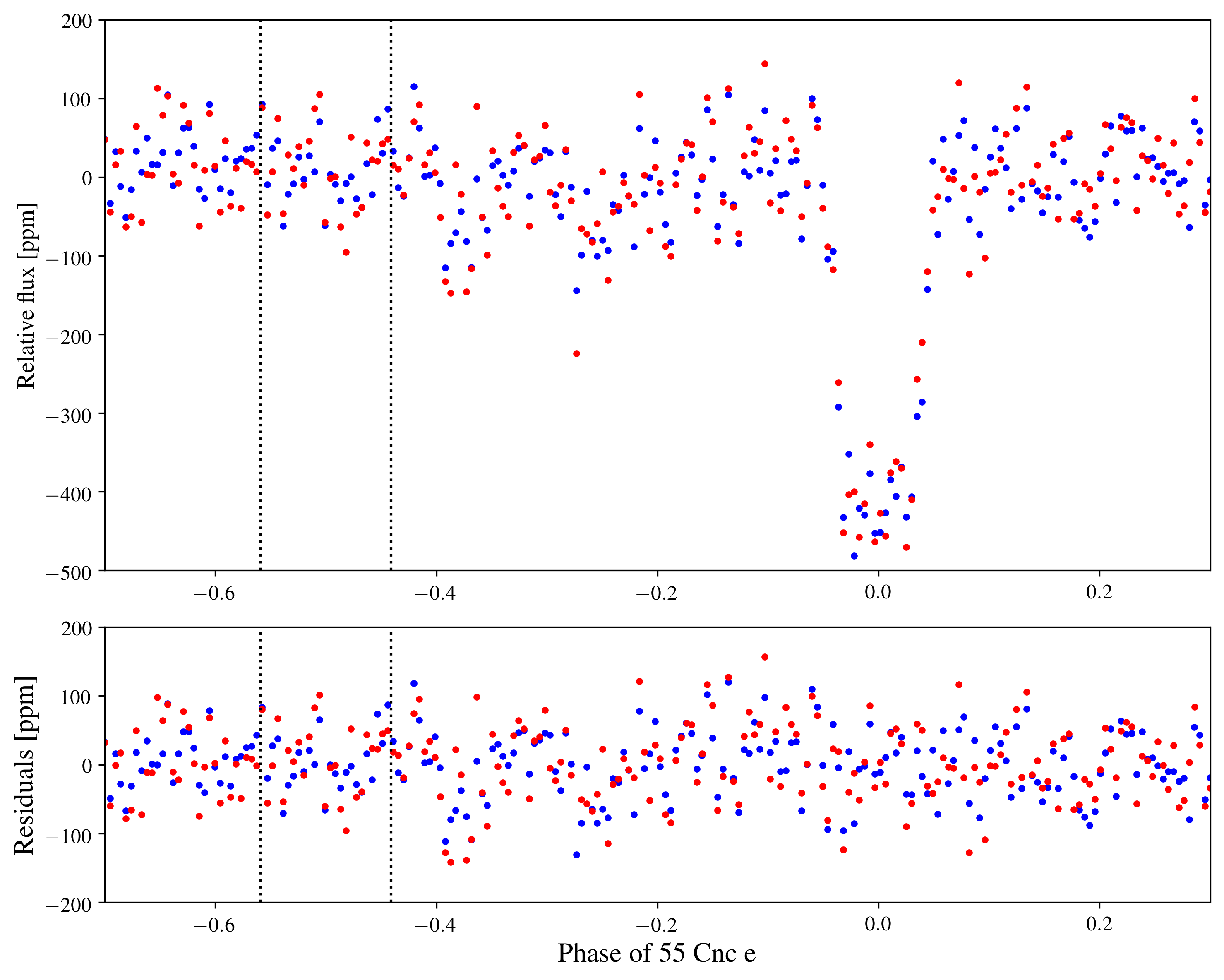}}
    \caption{Combine light curve phase folded at the planet orbital period and binned into $5$-min intervals (top) and associated residuals (bottom). The observed phase modulation has been removed using results given in Sec.~\ref{sec42} to compare only the transit shape. Observations detrended by the classical procedure are shown in blue and by the time-shift procedure in red. At the timescale of the planet orbital period, the light curves are globally identical (S/N $\sim$ 40, rms $\sim$ 1000 ppm). The phase range of the secondary eclipse are indicated by the vertical dotted lines.}
    \label{Fig_lcs_all}
\end{figure}

\begin{figure*}[hp!] \centering
    \resizebox{0.95\hsize}{!}{\includegraphics{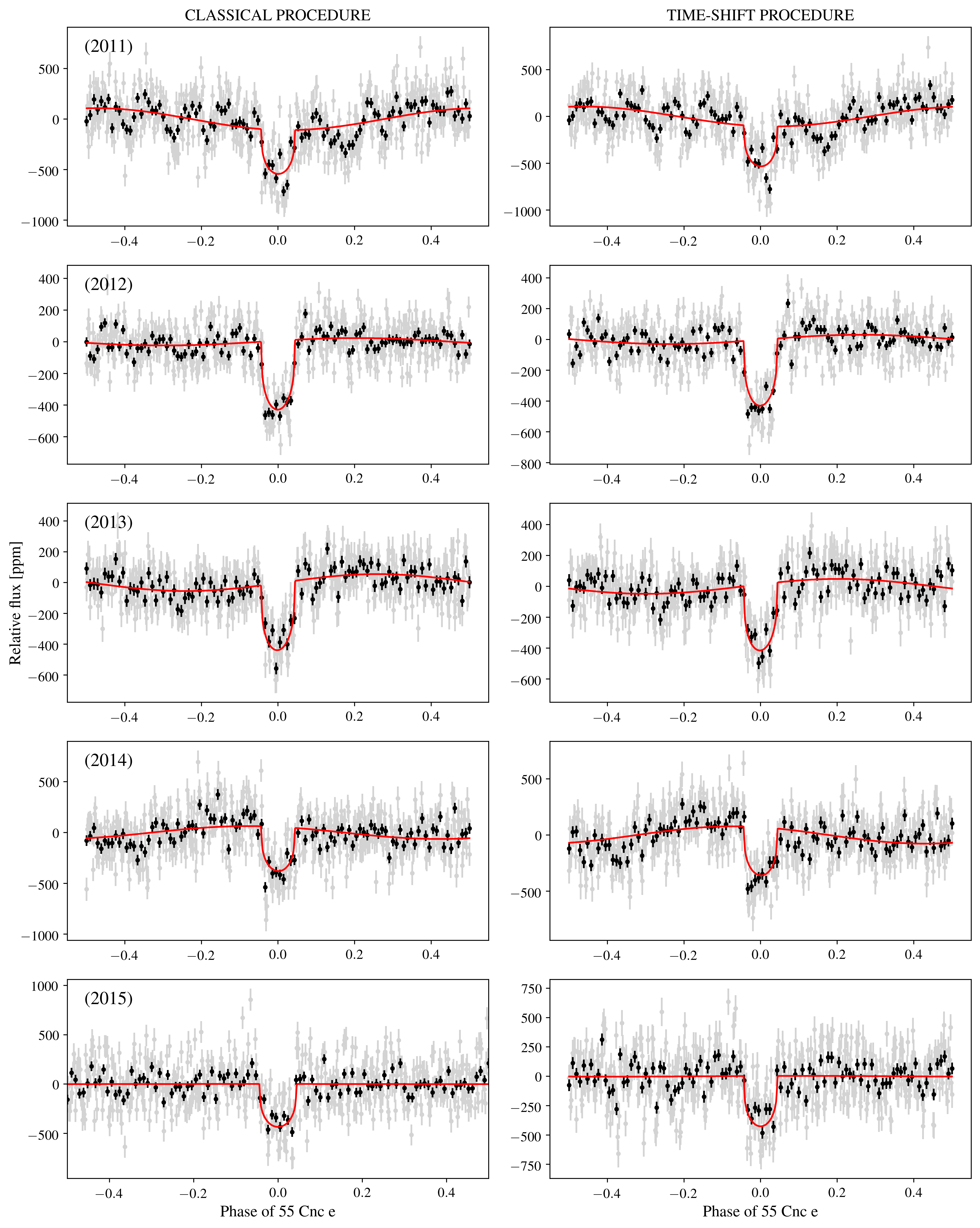}}
    \caption{Light curves phase folded at the planet orbital period and binned into $2$ (gray) and $10$-min (black) intervals. From top to bottom, we show data taken in 2011, 2012, 2013, 2014, and 2015. The first column shows  the final light curves obtained via the classical detrending procedure and the second column shows the light curves obtained via the time-shift detrending procedure. In these plots, the uncertainties are based on the original unscaled photometric uncertainties (in contrast to the scaled errors used in the MCMC analyses, see Sec.~\ref{sec44}). The best-fitting models are shown in red. 
}
    \label{Fig_lcs}
\end{figure*}

\begin{figure*}[ht!] \centering
    \resizebox{\hsize}{!}{\includegraphics{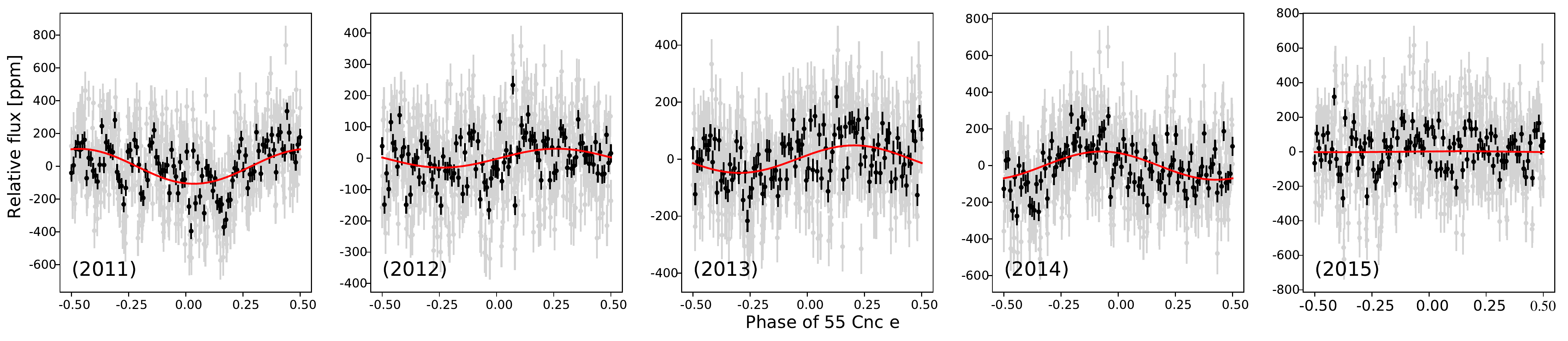}}
    \caption{Light curves detrended by the time-shift procedure, without transits, phase folded at the planet orbital period and binned into $2$ (gray) and $10$-min (black) intervals.}
    \label{Fig_modulation}
\end{figure*}

The middle box in Table~\ref{table0} lists the parameters involved in this new procedure as well as its performance in terms of rms and transit S/N. While the increase (resp. decrease) of the transit S/N (resp. rms) is not drastic, the benefit of this procedure can be seen in the comparison of the final light curves. 
The 2015 light curve phase folded at the spacecraft orbital period resulting from this time-shift procedure is shown in red in the bottom right panel of Fig.~\ref{Fig_steps}. We see that, even if it does not increase the transit S/N
(see last column of Table~\ref{table0}), the time-shift method significantly reduces the systematics induced by stray-light occurring at the timescale of the spacecraft orbit. 
The light curves phase folded at the planet orbital period are shown in the middle panel of Fig.~\ref{Fig_lcs}. 
Comparing these light curves with those obtained from the classical procedure (Fig.~\ref{Fig_lcs}, left column), we see a reduction of the remaining pattern modulated with the orbital period of the satellite ($\sim 1/10$ of the planet period). This effect is particularly remarkable for the 2011 dataset (top panels). However, as we see in Sec.~\ref{sec4}, this correction does not significantly affect the features occurring at the timescale of the planet orbital period ($> 10 P_{sat}$). At this timescale, the light curves detrended by both procedures remain comparable with a similar transit S/N when the light curves are combined (S/N of $40.2$, see Fig.~\ref{Fig_lcs_all}). A better visualization of the detected modulation in flux is shown in Fig.~\ref{Fig_modulation} for the light curves detrended by the time-shift procedure with the transit model removed. 

We finally note that the time-shift method could also be applied on the whole time series to correct the time delay between all orbits globally. However, as the shape of the stray-light pattern changes significantly over the course of several days, this would degrade the accuracy of the correction.

\section{Transit properties of 55 Cnc e}
\label{sec4}

To interpret our data, we performed a Markov Chain Monte Carlo (MCMC) analysis using both the light curves detrended via the classical method, with and without the time-shift correction. A detailed description of the MCMC scheme can be found in \citetads{2017A&A...606A..18L}. We used the \citetads{2002ApJ...580L.171M} 
algorithm to model transits and occultations, and the differential-evolution MCMC 
engine described in \citetads{2017AJ....153....3C}. As our datasets do not have 
a high enough S/N to fit for stellar limb darkening, we used a quadratic 
limb-darkening law with fixed parameters that have been derived for the MOST 
bandpass: $u_1= 0.648$ and $u_2=0.117$ (D14). To derive the transit, phase variation and secondary eclipse parameters, 
we carried out our analysis in three steps.

First, we fit for the transit period ($P$), epoch of mid-transit ($T_0$), planet-to-star radius ratio ($R_p/R_s$), impact parameter ($b$), and 
transit duration ($t_d$) to estimate precisely the planetary orbital period.
Second, we fixed the orbital period and proceeded to a second fit of the light curve, combining the transit model with a function modeling the variations in flux observed at the planetary period (see Fig.~\ref{Fig_lcs}). We added these variations to the transit model as sinusoidal functions of (fixed) planetary orbital frequency $f_s=1/P$: 
\begin{equation}
    F_{\rm mod}(t) = \boldsymbol{\alpha}_{\rm mod}~ \sin{(2\pi f_s t+ \boldsymbol{\phi}_{\rm mod})},
\label{eq_ec}
\end{equation}
where the vectors $\boldsymbol{\alpha}_{\rm mod}$ and $\boldsymbol{\phi}_{\rm mod}$ collect all the information about the amplitudes and orbital phases (relative to mid-transit) of the various datasets (simplified hereafter as $\{\alpha_i, \phi_i\}$ with $i \in [2011,2012,2013,2014,2015]$). In this analysis, we simultaneously estimated $14$ free parameters  ($4$ for the transit and $5\times2$ for the phase 
variation). Finally, we removed the best-fitting models of transit and phase 
modulation from the time series and fit the secondary eclipse depth.

\begin{figure*}[ht!] \centering
    \resizebox{\hsize}{!}{\includegraphics{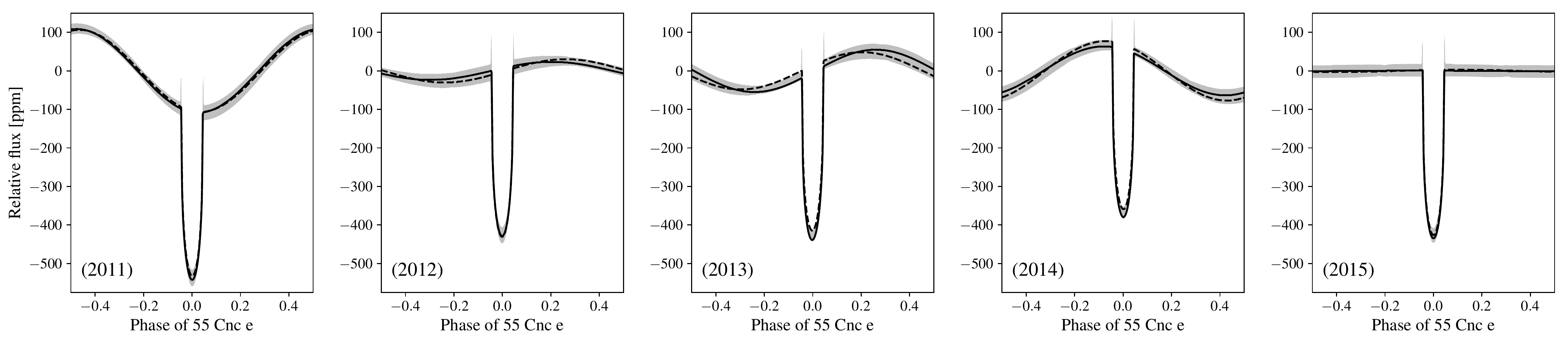}}
    \caption{Best-fitting models resulting from the MCMC analysis as described in Sec.~\ref{sec4}, using the light curves detrended by the classical (solid) and time-shift (dashed) procedures. For the latter only, we display the $1\sigma$ uncertainties (gray shaded area). These uncertainties have been evaluated using $10^4$ transit models generated using transit and phase modulation parameters randomly taken from the MCMC posteriors.
}
    \label{Fig_models}
\end{figure*}

Following the reassessed transit parameters of \citetads{2018A&A...619A...1B}, we assumed an eccentricity of zero for the orbit of the planet and added a prior on the impact parameter ($b=0.39\pm 0.03$) to help convergence.
The results are described below and the best-fitting models for transit and phase modulation for each year are compared in Fig.~\ref{Fig_models}. 

\subsection{Updated 55 Cnc e parameters}
\label{sec41}

With $143$ transit events, we precisely estimated the planetary orbital 
period of 55 Cnc e.
This value, found during the first MCMC analysis, is given in the first row of Table~\ref{table_tparam}. Results found on the time series detrended by the classical and time-shift methods are in complete $1\sigma$ agreement with each other as well as with the period extracted through velocity measurements \citepads{2018A&A...619A...1B}.
When estimating the flux modulation parameters, the orbital period has 
to be fixed to ensure the convergence of the second MCMC. The inferred 
transit and phase modulation parameters obtained during this second fit are given 
in Tables~\ref{table_tparam} and \ref{table_pv}, respectively. 
The associated marginal posteriors are reported in Fig.~\ref{Fig_posterior} and the best-fitting models are shown in Fig.~\ref{Fig_models}.
Once more, we find similar parameters from light curves detrended via the 
classical and via the time-shift methods.
Moreover, the parameters derived from the first and second MCMC agree to better than $1\sigma$.
The most significant difference is for the transit depth that slightly decreases ($< 5$ ppm) between the two runs because of the addition of the phase modulation in the second fit. 
All values for the transit parameters are in $1\sigma$ agreement with the 
values published by \citetads{2018A&A...619A...1B} and the previous results 
based on MOST data by W11 and D14. However, we note that we find slightly shallower transits (at approx. $1\sigma$) than W11 and D14. 

\subsection{Phase variations}
\label{sec42}

\begin{figure}[th!] 
    \resizebox{\hsize}{!}{\includegraphics{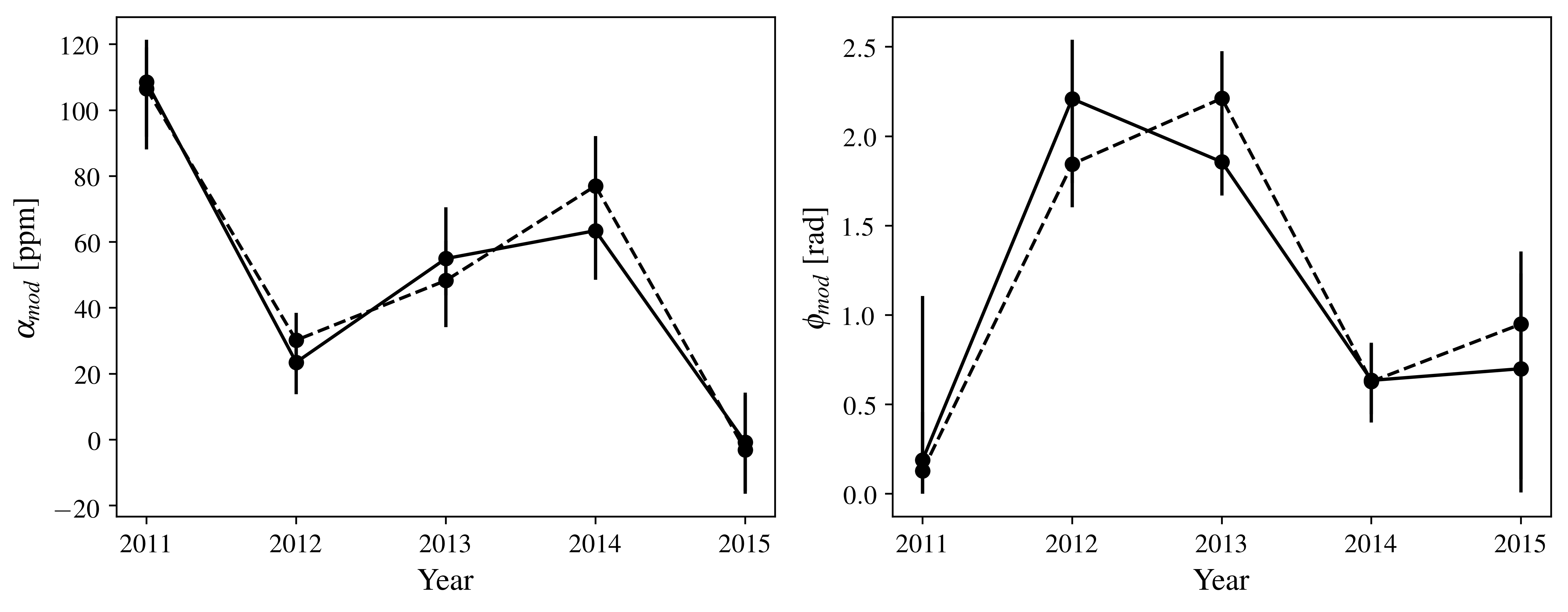}}
    \caption{Evolution of the amplitudes (left) and orbital phases (right) parameters of the modulation in flux observed at the planet orbital period (see Eq.~\ref{eq_ec}). The solid and dashed lines 
indicate the light curves detrended by the classical and time-shift procedures, 
respectively. }
    \label{Fig_PV}
\end{figure}

Contrary to previous analyses of MOST data (W11, D14), our phase curve model (see Eq. \eqref{eq_ec}) allows the flux maximum to be offset in time from the planetary occultation. 
Furthermore, we fit independent phase curve parameters for each year, thus probing temporal variability. 
Fig.~\ref{Fig_PV} illustrates the evolution of the phase modulation
from year to year. The phase variation, initially observed in the 2011 dataset by W11, is seen during most of the subsequent years, but its phase and amplitude change.
In 2011, the phase variation amplitude is the highest. In 2012, it seems to be 
attenuated in comparison to the 2011 dataset. In 2013 and 2014, it 
is present, although with a different phase and amplitude. 
We find no modulation in 2015, however this dataset is the noisiest and the 
phase variation might be masked by residual correlated noise (see rms values given in Table~\ref{table0}). 
Compared to W11, the derived amplitude of the phase modulation measured in the 
2011 dataset is smaller\footnote{Winn et al. found $\alpha_{2011}= 168 \pm 70$ 
ppm.} but agrees within $1\sigma$. 
When analyzing the light curves of 2011 and 2012 together, and fitting for a common phase modulation for both years, we find an amplitude of $\alpha_{2011+2012}= 24 
^{+8.7} _{-7.8}$ ppm that agrees with the results\footnote{Dragomir et al. 
found $\alpha_{2011+2012}= 34 ^{+12} _{-11}$ ppm.} of D14. However, as the 
modulation changes significantly with time (see Fig.~\ref{Fig_PV}), we argue 
that combining light curves from several years tends to attenuate the observed modulation.

We note that we attempted to probe shorter timescales of the phase curve variability by studying subsets of our five MOST datasets, but the quality of the data at hand are not sufficient to draw meaningful conclusions; i.e., the signal at the planet period  only significantly appears when a sufficiently large number of planet periods are phase folded.

\begin{table*}[!t] \centering
\caption{Inferred transit parameters for 55 Cnc e obtained using the light curves detrended by the classical (CM) and time-shift (TM) procedures and their $1\sigma$ uncertainties. The last column indicates the most recent values published in \citetads{2018A&A...619A...1B}.  Notes: All the $1\sigma$ uncertainties were derived using the distribution of the parameter posteriors. The impact parameter was completely determined by the input prior.}
{\setlength{\extrarowheight}{1pt}
\begin{tabular}{|c||ccccc||c|}
\cline{2-7} 
\multicolumn{1}{c|}{} &\bf Parameter & \bf Symbol & \bf Units & \bf Value (CM) & 
\bf Value (TM)  &  Bourrier et al. \\
\cline{2-7}  \hline
\cline{2-7} \hline
&\rule{0pt}{0.25\normalbaselineskip}          & & & &  &  \\
&Orbital period   & $P$                  & days         & $0.73654530 
^{+6.5\times10^{-7}} _{-9.5\times10^{-7}}$ & $0.73654504 ^{+7.6\times10^{-7}} 
_{-9\times10^{-7}}$   &  $0.73654737^{+1.3\times10^{-6}} _{-1.44\times10^{-6}}$\\
&\rule{0pt}{0.25\normalbaselineskip}          & & & &  &  \\
&Transit epoch    &  $T_0 - 2~451~545$   & BJD          & $4417.0720 
^{+6.1\times10^{-4}} _{-3.6\times10^{-4}}$ & $4417.0719 ^{+6.6\times10^{-4}} 
_{-5.0\times10^{-4}}$  &  $4417.0712 
^{+1.4\times10^{-3}} _{-1.4\times10^{-3}}$ \\
&\rule{0pt}{0.25\normalbaselineskip}          & & & &  &  \\
&Planet-to-star radius ratio & $R_p/R_s$ &   & $0.01874 
^{+3.9\times10^{-4}} _{-2.9\times10^{-4}}$ & $0.01860 ^{+3.2\times10^{-4}} 
_{-4.0\times10^{-4}}$  &  $0.0182 
^{+2\times10^{-4}} _{-2\times10^{-4}}$\\
&\rule{0pt}{0.25\normalbaselineskip}          & & & & &   \\
&Impact parameter  & $b$                 &  & $0.3917 ^{+0.029} 
_{-0.033}$ & $0.3954 ^{+0.027} _{-0.035}$ & $0.39 ^{+0.03} 
_{-0.03}$  \\
&\rule{0pt}{0.25\normalbaselineskip}          & & & &  &  \\
\rot{\rlap{\bf ~Measured parameters}}   
&Transit duration  & $t_d$               & days        & $0.0648 
^{+1.1\times10^{-3}} _{-0.9\times10^{-3}}$ & $0.0654 ^{+0.9\times10^{-3}} 
_{-1.4\times10^{-3}}$  &  $0.0634 ^{+3.7\times10^{-4}} 
_{-3.7\times10^{-4}}$\\  
&\rule{0pt}{0.25\normalbaselineskip}          & & & & & \\
&Eclipse depth  & $\delta_{ecl}$        & ppm & $- 9.99 ^{+7.5} _{-10.6}$ & $- 
12.88 ^{+8.9} _{-7.6}$  &  \\
&\rule{0pt}{0.25\normalbaselineskip}          & & & &  &  \\
\cline{2-7}  \hline \hline
&\rule{0pt}{0.25\normalbaselineskip}          & & & &  &  \\
&Transit depth  & $(R_p/R_s)^2$ & ppm  & $351.33 ^{+11.16} _{-14.68}$ & 
$346.11 ^{+11.99} _{-14.90}$  & $331.24 ^{+7.28} _{-7.28}$  \\
&\rule{0pt}{0.25\normalbaselineskip}          & & & & &   \\
&Planetary radius   & $R_p$                  & $R_\oplus$         & $1.8865 
^{+0.045} _{-0.035}$ &  $1.90 ^{+0.037} _{-0.047}$  &  $1.875 
^{+0.029} _{-0.029}$ \\
&\rule{0pt}{0.25\normalbaselineskip}          & & & &  &  \\
&Scaled semimajor axis   & $a/R_s$                  &          
& $3.4746 ^{+6.9\times10^{-2}} _{-7.2\times10^{-2}}$ &   $3.5046 
^{+6.4\times10^{-2}} _{-9.4\times10^{-2}}$ & $3.52 ^{+1.0\times10^{-2}} _{-1.0\times10^{-2}}$ \\
&\rule{0pt}{0.25\normalbaselineskip}          & & & & &   \\
&Semimajor axis   & $a$                  & AU         & $0.01521 
^{+3.7\times10^{-4}} _{-3.4\times10^{-4}}$ & $0.01528 ^{+4.1\times10^{-4}} 
_{-3.5\times10^{-4}}$  &   $0.01544 
^{+5.0\times10^{-5}} _{-5.0\times10^{-5}}$\\
&\rule{0pt}{0.25\normalbaselineskip}          & & & &  &  \\
&Orbital inclination   & $i$                  & deg         & $83.56 
^{+0.61} _{-0.62}$ &  $83.72 ^{+0.49} _{-0.74}$ &   $83.59 ^{+0.47} _{-0.44}$ \\
\rot{\rlap{\bf ~~~~~Derived parameters}}
&\rule{0pt}{0.25\normalbaselineskip}          & & & & &   \\
\cline{2-7}  \hline 
\end{tabular} }
\label{table_tparam}
\end{table*}

\subsection{Secondary eclipse}
To search for the secondary eclipse of 55 Cnc e, we first divided the data by our best-fitting transit and flux  modulation models. Then, we performed a final MCMC 
analysis of the light curve residuals of all observations. 
Previous analyses from D14 did not detect the secondary eclipse, and these authors estimated the depth of this eclipse to be $-1^{+18} _{-22}$ ppm (using the MOST data of 2011 and 2012).
Expecting an extremely shallow signature of the planetary eclipse and to ensure obtaining a representative posterior distribution, we allowed the depth parameter to take negative values in our MCMC analysis (no physical meaning).

We find an eclipse depth of $\delta_{ecl} = - 12.88 ^{+8.9}_{-7.6} $ ppm. Consequently, we do not detect any optical signature of the secondary eclipse of 55 Cnc e, but we can place a $2\sigma$ limit of $16$~ppm on its depth using the posterior distribution. Fig.~\ref{Fig_lcs_all} shows the multiyear light curve phase folded on the period of 55 Cnc e, where the phase range of the secondary eclipse is indicated by dotted lines.

We estimated the upper limit on the geometric albedo ($A_g$) of 55 Cnc e using the following equation \citepads{Row08}:
\begin{equation}
\label{eqn:albedo}
\delta_{ecl} = A_g\left(\frac{R_p}{a}\right)^2.
\end{equation}

Given the very close orbit of 55 Cnc e, we compared its predicted thermal contribution with our upper limit on the eclipse depth using our $R_p/R_s$ value, a stellar effective temperature of $5172 \pm 18$ K \citepads{2017ApJ...836...77Y}, and two values for the planet temperature: $2300$ K, which is the highest value of the predicted equilibrium temperature (with zero albedo and zero heat redistribution; \citeads{Cro12}); and $2700$ K, which is the hemisphere-averaged value measured by \citetads{Demory_2016_nature}. The thermal contribution is $0.9$ and $4$ ppm, for each of those two cases, respectively. In this section we derived $2\sigma$ albedo limits for both cases, but we used and refered to the value corresponding to the measured temperature (2700 K) throughout the remaining sections of this paper.

We first subtracted the thermal contribution from the $2\sigma$ limit on $\delta_{ecl}$, and used the resulting value in equation \ref{eqn:albedo}. We then derived a $2\sigma$ lower limit on $R_p/a$ from the corresponding lower limit on $R_p/R_s$ from the second-to-last column of Table~\ref{table_tparam}, and from the corresponding upper limit on $a/R_s$ from \citet[][as their value is in agreement with and more precise than ours]{2018A&A...619A...1B}. We thus obtained $2\sigma$ upper limits on $A_g$ of $0.59$ and $0.47$, assuming planet temperatures of 2300 and 2700 K, respectively. 

\begin{table*}[!t] \centering
\caption{Inferred parameters of the modulation in flux observed at the planet orbital period. Results obtained using the light curves detrended by the classical and time-shift procedures are shown at left and right, respectively.}
\setlength\tabcolsep{10.5pt}
\renewcommand{\arraystretch}{0.8}
\begin{tabular}{||c|c|c||} 
\multicolumn{3}{c}{\bf Classic}\\
\multicolumn{3}{c}{ }\\
\hline
Year & \bf Amplitude ($\boldsymbol{\alpha}_{\rm mod}$) & \bf  Phase ($\boldsymbol{\phi}_{\rm mod}$) \\
\hline \hline
\rule{0pt}{0.5\normalbaselineskip}  & &  \\
$2011$  & $108.54 ^{+12.8} _{-16.6}$ ppm & $0.19 ^{+0.27} _{-0.19}$ rad \\
\rule{0pt}{0.5\normalbaselineskip}  & &  \\
$2012$  & $23.49 ^{+8.2 } _{-9.7}$ ppm & $2.20 ^{+0.33} _{-0.47}$ rad \\
\rule{0pt}{0.5\normalbaselineskip}  & &  \\
$2013$  & $54.96 ^{+15.6} _{-11.2}$ ppm & $1.85 ^{+0.27} _{-0.19}$ rad \\
\rule{0pt}{0.5\normalbaselineskip}  & &  \\
$2014$  & $63.43 ^{+14.1} _{-14.9}$ ppm & $3.98  ^{+0.21} _{-0.26}$ rad \\
\rule{0pt}{0.5\normalbaselineskip}  & &  \\
$2015$  & $-0.77 ^{+14.3  } _{-14.7}$ ppm & $0.70 ^{+0.65} _{-0.62}$ rad \\
\rule{0pt}{0.5\normalbaselineskip}  & &  \\
\hline
\end{tabular}
\hspace{2cm}
\begin{tabular}{||c|c|c||} 
\multicolumn{3}{c}{\bf Time-shift}\\
\multicolumn{3}{c}{ }\\
\hline
Year & \bf Amplitude ($\boldsymbol{\alpha}_{\rm mod}$) & \bf  Phase ($\boldsymbol{\phi}_{\rm mod}$) \\
\hline \hline
\rule{0pt}{0.5\normalbaselineskip}  & &  \\
$2011$  & $106.52 ^{+12.5} _{-18.4}$ ppm & $0.13 ^{+0.26} _{-0.13}$ rad \\
\rule{0pt}{0.5\normalbaselineskip}  & &  \\
$2012$  & $30.19 ^{+8.3 } _{-9.5}$ ppm & $1.84 ^{+0.36} _{-0.24}$ rad \\
\rule{0pt}{0.5\normalbaselineskip}  & &  \\
$2013$  & $48.3 ^{+12.1} _{-14.2}$ ppm & $2.21 ^{+0.26} _{-0.29}$ rad \\
\rule{0pt}{0.5\normalbaselineskip}  & &  \\
$2014$  & $77.01 ^{+15.1} _{-15.3}$ ppm & $3.94  ^{+0.20} _{-0.18}$ rad \\
\rule{0pt}{0.5\normalbaselineskip}  & &  \\
$2015$  & $-3.05 ^{+17.3  } _{-13.4}$ ppm & $0.94 ^{+0.28} _{-0.30}$ rad \\
\rule{0pt}{0.5\normalbaselineskip}  & &  \\
\hline
\end{tabular}
\label{table_pv}
\end{table*}


\subsection{Uncertainties of the inferred parameters}
\label{sec44}

To validate the uncertainties on both the phase modulation parameters and the 
eclipse depth, we performed injection tests as done in W11.
These tests consist in injecting a synthetic signal (either the secondary eclipse or the phase modulation) in the light curve residuals and performing an MCMC analysis of these synthetic light curves to estimate the parameters of interest. For both cases, we used the 2013 data only to save on computation time and performed $1000$ individual injections. We used an eclipse depth of $\delta_{ecl} = 5$ ppm and a phase modulation amplitude of $\alpha_{13} = 45.63$ ppm. When injecting secondary eclipses, the time at mid-eclipse  is randomly chosen between $[t_0-t_d/2,t_0+P-t_d/2]$. 
When injecting a phase modulation, the phase parameter $\Phi_{13}$ is randomly chosen between $[-2\pi,2\pi]$.

\begin{figure}[h!] \centering
\resizebox{0.975\hsize}{!}{\includegraphics{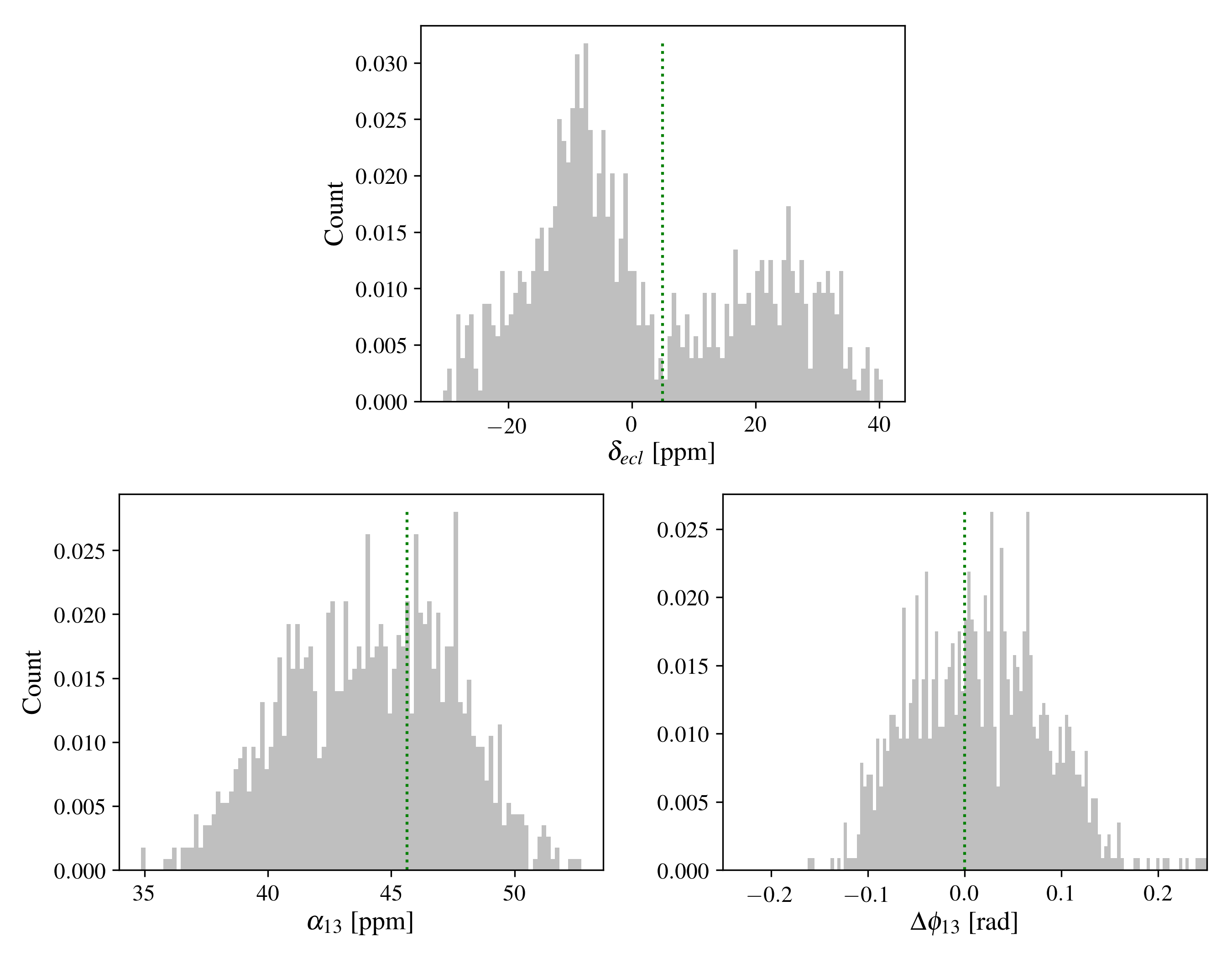}}
\caption{Normalized distribution of the best-fitting values obtained from 
injection tests performed for the planetary eclipse (top) and for the phase 
modulation (bottom). The synthetic eclipse depth has been set to $5$ ppm and the phase modulation amplitude to $\alpha_{13}=48.3$ ppm. In the bottom right panel, we show the difference between the inputs and estimates of the phase parameter ($\Delta \Phi_{13}$) as the input phase value is randomly changing from one test to 
another.}
\label{Fig_boot}
\end{figure}

Fig.~\ref{Fig_boot} shows the distribution of the best-fitting values for the 
recovered eclipse depth (top) and the parameters $\alpha_{13}$ and $\Phi_{13}$ 
(bottom). For the secondary eclipse depth, we find a bimodal distribution around the true injected value (green dashed line). This distribution is centered around the true value, but is clearly not Gaussian. This illustrates the influence of 
the correlated noise still present in the reduced light curves. Depending on the 
injected secondary eclipse timing, its depth is over --- or under --- estimated as a consequence of the average level of correlated noise at this orbital phase.
From these tests, we estimate a realistic uncertainty on the occultation depth of up to $35$ ppm (measured as the maximum dispersion around the true value of $5$ ppm).

For the phase modulation, the distribution of the retrieved values for amplitude and phase are centered around the true values. 
The widths of the two distributions are $7$~ppm and $0.2$~rad for amplitude and phase, respectively.  
These results show that the impact of the remaining correlated noise is less significant on timescales longer than the satellite orbital period; i.e., they do not affect the modulation in flux related to the planet orbit, for which we also have many more photons.

Beside the injection tests, we also tested whether reliable uncertainties can be obtained by scaling the input errors.
The MCMC code used in this work uses the $\chi^2$ as its merit function, however this implicitly assumes white Gaussian noise. 
As correlated noise persists in the data even after the corrections described in Section \ref{sec31}, this would lead to largely underestimated errors on the derived parameters. 
Therefore, we scale our photometric errors. We use the $\beta_r$ factor that compares the ratio of standard deviations evaluated on the binned and raw residuals \citepads{2008ApJ...683.1076W, 2010A&A...511A...3G}. 
We tested bin intervals between $5$ and $20$ minutes and scaled the error bars by the highest value found. 
We find that a relatively large factor of $\beta_r=2.262$ allows us to obtain similar uncertainties on the phase curve parameters as those derived using injection tests.

The uncertainties given in Tables~\ref{table_tparam} and \ref{table_pv} were derived using this $\beta_r$ method. We note that for the secondary eclipse search, we used the errors scaled during the second MCMC analysis.

\begin{figure*}[th!] \centering
    \resizebox{\hsize}{!}{\includegraphics{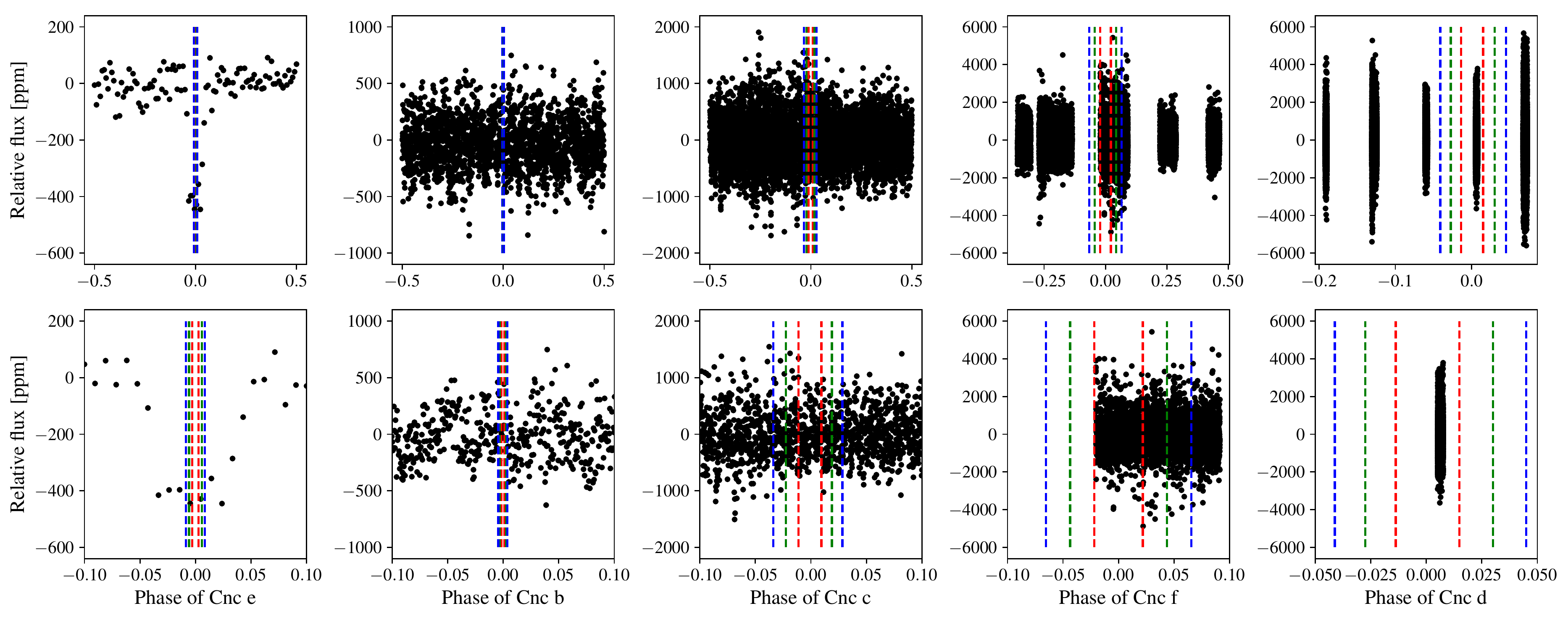}}
    \caption{From left to right: Light curves binned into $5$-min intervals, phase folded at the orbital period of planet e, b, c, f, and d. The vertical lines indicate the $1$, $2,$ and $3\sigma$ uncertainties of the time of mid-transit (red, green, and blue, respectively). The bottom panels are zoomed-in views of the mid-time transit event for each planet.}
    \label{Fig_pls}
\end{figure*}

\section{Search for transits of 55 Cnc b, c, f, and d}
\label{sec:other}

There are four other known planets in the 55 Cnc system that have been detected via radial velocity measurements \citepads{Fischer1, Dawson}. Given that the innermost one, i.e., planet e, transits the host star, there is a non-negligible probability that one or more of the others may transit as well. Their masses and orbital periods span wide ranges \citepads{Dawson}, forming an extrasolar multiple planet system that is moderately similar to our solar system. 55 Cnc f holds particular interest because it spends about $74$\% of its eccentric orbit within the habitable zone of the system \citepads{Braun3}; we note that this statement may no longer be true as the recent re-evaluation of the eccentricity is now consistent with zero (\citeads{2018A&A...619A...1B}).  

Using parameters given in Table 3 of \citetads{2018A&A...619A...1B}, we computed the predicted transit times during our MOST observations for the five innermost planets. Fig.~\ref{Fig_pls} shows the MOST data phase folded at the orbital period of the planets with the $1$, $2,$ and $3\sigma$ uncertainties of their time at mid-transit. 
When using the MCMC routines described above to search for transits of each of these planets, we come up empty. However, assuming the innermost planets b and c were transiting, we can give an upper limits on their planetary radii. 
Considering $b=0.5$ as the nominal value (because the orbital inclination of each of these planets' orbit is unknown), we predict an upper limit of $2.15~R_\oplus$ for planet 55 Cnc b and $2.56~R_\oplus$ for planet 55 Cnc c (assuming a limiting S/N of $40$ as planet e).

\section{Discussion}
\label{sec:disc}

\subsection{Albedo of 55 Cnc e}
While we do not detect a secondary eclipse of 55 Cnc e, we can consider the $0.47$ upper limit on its albedo in the context of other albedo measurements in the literature. Thanks to {\it Kepler} and {\it K2}, it has been determined that hot Jupiters are typically dark and likely cloudless (i.e., $A_g < 0.2$; \citealt{Esteves2015,Angerhausen2015}). A few hot Jupiters have $A_g$ values between 0.2 and 0.35 \citep{Esteves2015,Angerhausen2015,Demory11b}, indicating that a small fraction of these planets have clouds.

However, 55 Cnc e should be compared to planets of its own size. Albedo measurements of the Kepler close-in super-Earth and Neptune sample as a whole have revealed that these worlds are somewhat less dark than hot Jupiters, but still have $A_g$ values generally below 0.3 \citep{Dem14, She17, Jan18}. One exception is the hot rocky exoplanet Kepler-10b \citep{Bat11}. \cite{Rou11} suggested ThO$_2$ particles dispersed in Al$_2$O$_3$-CaO lava as a possible origin for the relatively high (0.32) geometric albedo of the planet.

If the surface of 55 Cnc e is molten \citep{Demory_2016_nature} and assuming that at visible wavelengths we can view all the way to its surface, \citetads{2011ApJ...740...61K} predicted a secondary eclipse depth of 20 ppm using an $A_g$ value of 0.6. However, recent preliminary laboratory measurements of specular reflection from molten lava and quenched glass (a product of rapidly cooled lava) suggest an upper limit on the albedo of such a planetary surface of 0.1 (Zahra Essack, MIT, private communication). While we note that the albedo of molten rock depends on the composition of the mantle of the planet, which is unknown for 55 Cnc e, a molten surface unobstructed by an atmosphere (at the wavelengths probed by the MOST data) remains possible within our $A_g$ upper limit.

While our constraints on the geometric albedo of 55 Cnc e do not rule out any of the most likely atmospheric composition models (e.g., CO, CO$_2$, H$_2$O, N$_2$, O$_2$, HCN; \citealt{Ang17, Mig19}), we can provide a first test of the model proposed by \cite{Tam18} to explain the previously observed IR secondary eclipse variability \citep{Demory_2016}. \cite{Tam18} found that refractory particulates produced by volcanic activity, at times lofted high in the atmosphere, could obscure the surface of the planet and potentially explain the decrease in observed thermal emission. According to this model, 55 Cnc e would have a higher albedo (between 0.4 and 1.0) when its surface is obscured by the refractory particulates. In this scenario, this stage of the variability (e.g., when Spitzer $4.5$ $\mu$m secondary eclipse depth was at its lowest) was observed in 2012. The photometric precision of the MOST 2012 light curve alone is not sufficient to set a meaningful constraint on $A_g$ for that year. Nevertheless, our global constraint of $A_g < 0.47$ (averaged over all five MOST light curves) rules out most of the 0.4 - 1.0 range, and tentatively suggests that the model proposed by \cite{Tam18} may not explain the observed 4.5 $\mu$m secondary eclipse variability.
Ultimately, high-precision observations with the upcoming CHaracterising ExOPlanet Satellite (CHEOPS) space telescope \citepads{2013EPJWC..4703005B} would provide improved constraints on, and possibly an actual determination of, the albedo of 55 Cnc e.

\subsection{Possible origins of the phase modulations}
In this study, we analyzed five sequences of MOST data of 55~Cnc~e obtained between 2011 and 2015 and searched for photometric modulation in phase with the period of planet~e. We assumed a modulation that was periodic and estimated its parameters (amplitude and phase) for the five individual light curves. We detected a phase modulation and find that it is variable from year to year. Intriguingly, the amplitude of this modulation in flux is too large to be due to scattered light from the planet. We note that our measurements of the phase amplitude are also higher than the predictions by \citetads{2011ApJ...740...61K} who assumed various scenarios for the planet; the most optimistic is that of a lava world with a modulation amplitude $<20$ ppm. Consequently, the observed modulation must have a different origin.

\subsubsection{Instrumental origin}
First, it has been proposed by W11 that the variation may be due to instrumental noise. To verify this, we analyzed the light curve of the nearby star 53 Cnc, a bright giant star falling on the MOST CCD during the observations of 55 Cnc. We applied the same detrending procedures to the light curves of 53 Cnc and do not find any flux modulation at the timescale of the orbital period of 55 Cnc e. While we are pretty confident that the modulation observed in the 55 Cnc is not due to instrumental artifacts, this has to been confirmed by other long-term observations in the optical wavelength range. Assuming this modulation is indeed related to the 55 Cnc system, we can invoke multiple hypothetical physical scenarios to explain its origin. 

\subsubsection{Stellar variability }
We considered the possibility of variability of the star as the source of the phase modulation signal. The rotation period of 55 Cnc A is $\sim40$ days and the lifetime of star spots on solar-type stars ranges from 10 to 350 days \citep{Namekata19}. These timescales are too long for spot-related stellar variability to be the source of the signal. We also considered stellar pulsations, but in Sun-like stars p-mode oscillations have periods of just 5-10 minutes \citep{DiMauro2016}, which is much too short to be the source of the observed modulations at the period of 55 Cnc e.
We conclude that neither star spots nor stellar pulsations can explain the 55 Cnc e phase modulation signal.

\subsubsection{Star-planet interaction }
\begin{figure}[h!]\centering
    \resizebox{\hsize}{!}{\includegraphics{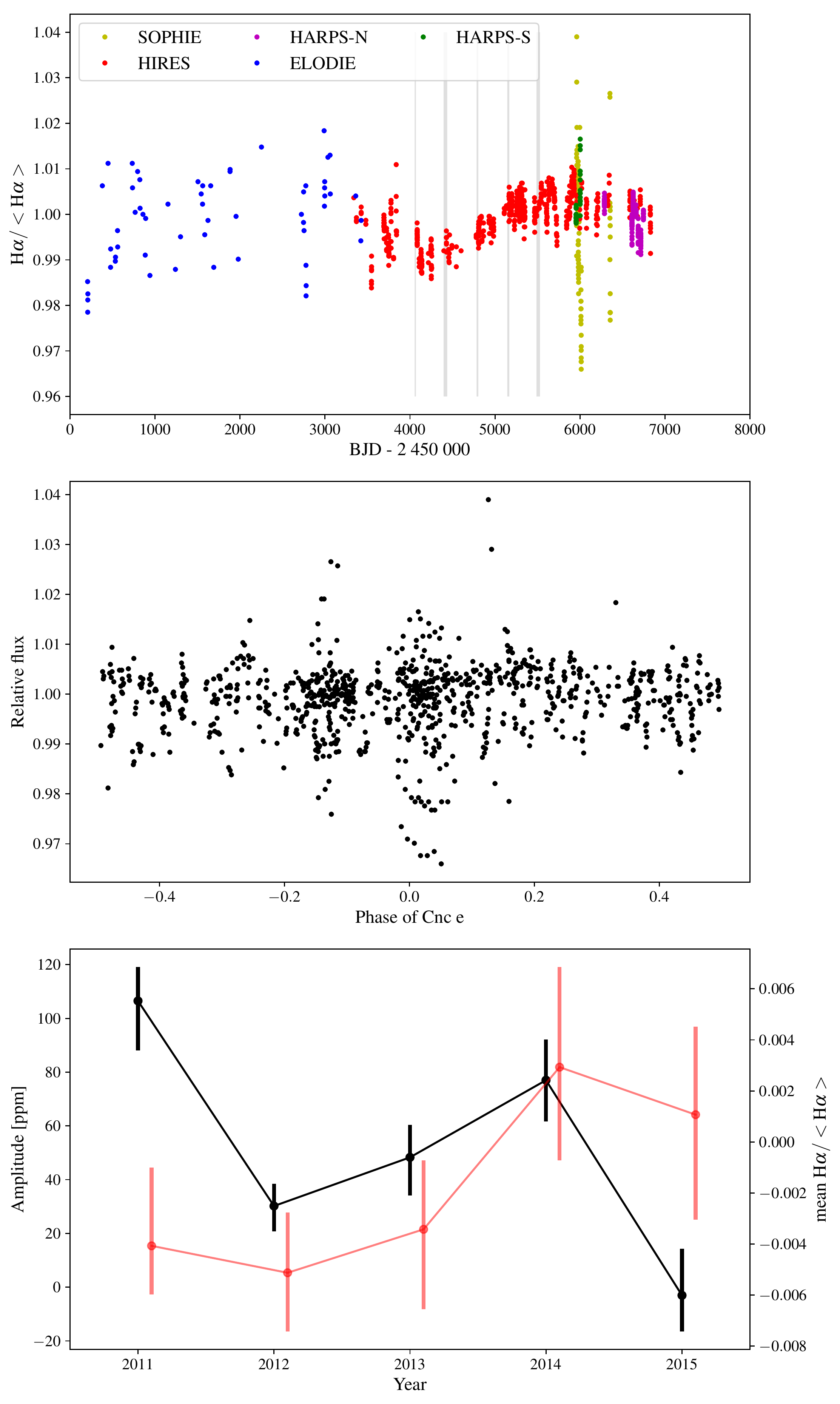}}
    \caption{Top: H-$\alpha$ activity index of the 55 Cnc A star (see Fig. 2 of \citeads{2018A&A...619A...1B}, second panel) and MOST observations dates (gray). Middle: Combined H-$\alpha$ data phase folded at the planet orbital period. Bottom: Amplitudes of the modulation (see Table.~\ref{table_pv}) measured on MOST observations (black, left y-axis) and mean values of the H-$\alpha$ stellar activity indicator (red, right y-axis) evaluated around dates close to the MOST observations. The error bars on the mean H-$\alpha$ values are taken as the minimum and maximum values of the index in the considered year. We note the H-$\alpha$ values have been slightly shifted in time for visibility. We observe the increase/decrease of the modulation in phase with the increase/decrease of the stellar activity indicator.}
    \label{Fig12_Halpha}
\end{figure}

As proposed by W11, this signal can be the signature of an interaction between the host star and planet e. 
For example, from ultraviolet observations \citetads{2018A&A...615A.117B} proposed that the interaction of the planet with the stellar magnetic field inside the corona might trigger coronal rain. The material would be accreted onto the star along the magnetic field lines, and cool and emit at optical wavelengths (see discussion in Sec. 4.3 of \citeads{2018A&A...615A.117B}). In this scenario, the variability in the observed modulation would be due to the variability of the stellar corona with time and to exchanges of variable amounts of material over time. 
To validate this scenario, we looked for modulations of the H-$\alpha$ stellar activity indicator obtained over the last $\approx$20 years \citep{2018A&A...619A...1B} in phase with the planetary orbital period, but without success (see Fig.~\ref{Fig12_Halpha}). This may be because the sampling and time coverage may be too poor to access the presence of such a modulation at the level allowed by the data quality. However, it is also possible that the signal is smeared out by the variation in the phase offsets we observe in the modulation from year to year, which is on timescales much shorter than the whole data coverage, even possibly shorter than the stellar rotation period. For this reason, we considered the subseries of stellar activity indicator measurements that are temporally close to the MOST observations. Figure~\ref{Fig12_Halpha} shows that there is indeed a similarity in the temporal behavior between the activity index and the amplitude of the flux modulation.

A supporting argument for the presence of star-planet interactions comes from the results of \citet{Folsom-inpress-a}, who employed spectropolarimetry to derive the strength and geometry of the stellar magnetic field. Their 3D stellar wind modeling based on the surface magnetic field map indicated that the orbit of the planet lies entirely inside the Alfv{\'e}n surface of the stellar wind. This implies that the star and the planet are magnetically connected and that the planet could influence the stellar wind such that the interaction can reach the stellar surface. If this scenario is correct, the observed flux modulation could be evidence of plasma exchange through reconnecting magnetic field lines of the star and planet.

The interaction could then lead to the formation of a spot on the stellar surface that rotates in phase with the planetary orbital motion, hence unrelated to the stellar rotation period \citepads{Shkolnik_2003, Strugarek_2015}. Such a scenario has already been proposed to explain the optical flux modulation observed for the $\tau$ Boo \citepads{Walker_2008} and CoRoT-2 \citepads{Pagano_2009} systems, as well as in X-rays for HD 17156 \citepads{Maggio_2015}. If a spot is present, calculations done for hot Jupiters have shown that the activity should be phased with the planetary orbital period and present a phase offset \citepads{Lanza_2012} similar to what was recently detected by \citetads{Cauley_2018}. Assuming this argument also holds for systems hosting lower mass planets and the quality and sampling of the data is high enough, the interaction should be detectable by analyzing the temporal behavior of magnetic activity indicators \citepads{Strugarek_2019}. Our results shown in Fig.~\ref{Fig12_Halpha} suggest that the interaction may indeed be detectable with simultaneous photometric and spectroscopic high-quality observations, the former covering the whole planet orbital period and the latter covering multiple activity indicators.

\subsubsection{Transiting circumstellar dust torus }

Alternatively, the phase modulation in flux could be the signature of a transiting circumstellar dust torus. This dust cloud could also exchange material with the planet itself, which might originate from volcanism at the surface of 55 Cnc e (analogous to Io’s cold plasma torus; \citeads{2003GeoRL..30.2101K}).
 The density and optical depth of the torus changes over time. Thus, both the transit depth and flux modulation vary (see Fig.~\ref{Fig15_Vincent}). This scenario would be in line with the variability observed in the IR secondary eclipse depth \citepads{Demory_2016,Tam18}.
Moreover, if the material of this torus indeed originates from the planet, it could be alimented from material (typically dust) inside the planetary Roche lobe ($\sim 4.94-5.22~R_\oplus$), which itself is highly irradiated and subject to intense tidal forces. In this paradigm, the proportion of scattered light also depends on the spatial distribution of the dust inside the Roche lobe and the position of the planet along its orbit. If the distribution of scattering material material evolved, the phase curve would as well.

\begin{figure*}[th!]\centering
    \resizebox{0.9\hsize}{!}{\includegraphics{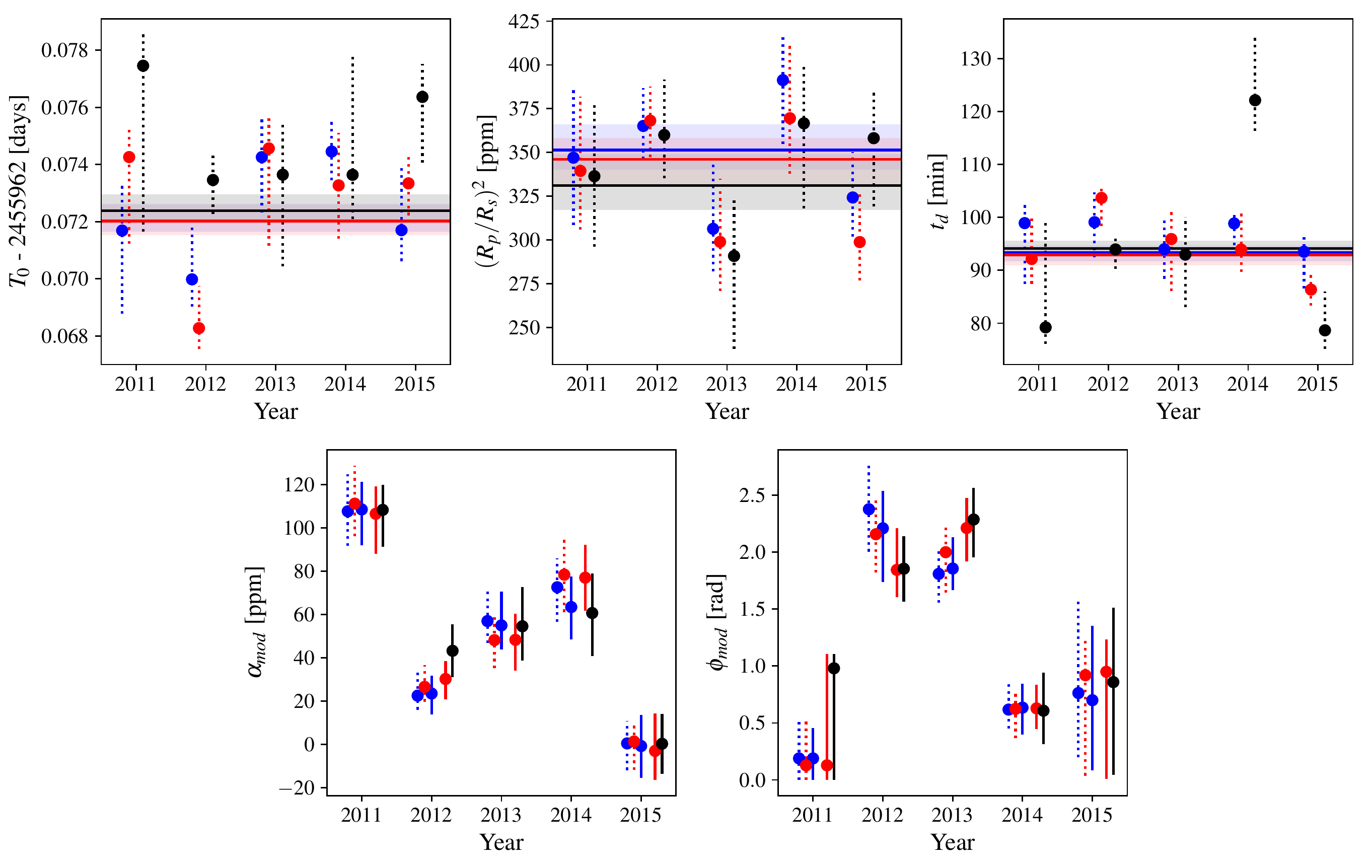}}
    \caption{From left to right: Estimated parameters of the 55 Cnc e transit ($T_0$, $(R_p/R_s)^2$, $t_d$) and of the modulation in flux ($\alpha_{\rm mod}$ and $\phi_{\rm mod}$). 
    In all panels, dotted points indicate the best-fitting values derived from each of the $5$ MOST data analyzed separately using the classical and time-shift detrending procedures and values obtained using the light curve not corrected for the stray-light systematics (i.e., resulting only from the pre-whitening step; see Sec.~\ref{sec31}). 
    In the three first panels, the horizontal solid lines represent the values derived from the combined light curves (see Table~\ref{table_tparam} and Sec.~\ref{sec41}) with their $1~\sigma$ uncertainties (shaded area).  
    In the two last panels, the solid points represent the modulation parameters derived during the analysis of these combined light curves (see Table~\ref{table_pv} and Sec.~\ref{sec42}). }
    \label{Fig8_params}
\end{figure*}

\begin{figure}[h!]\centering
    \resizebox{\hsize}{!}{\includegraphics{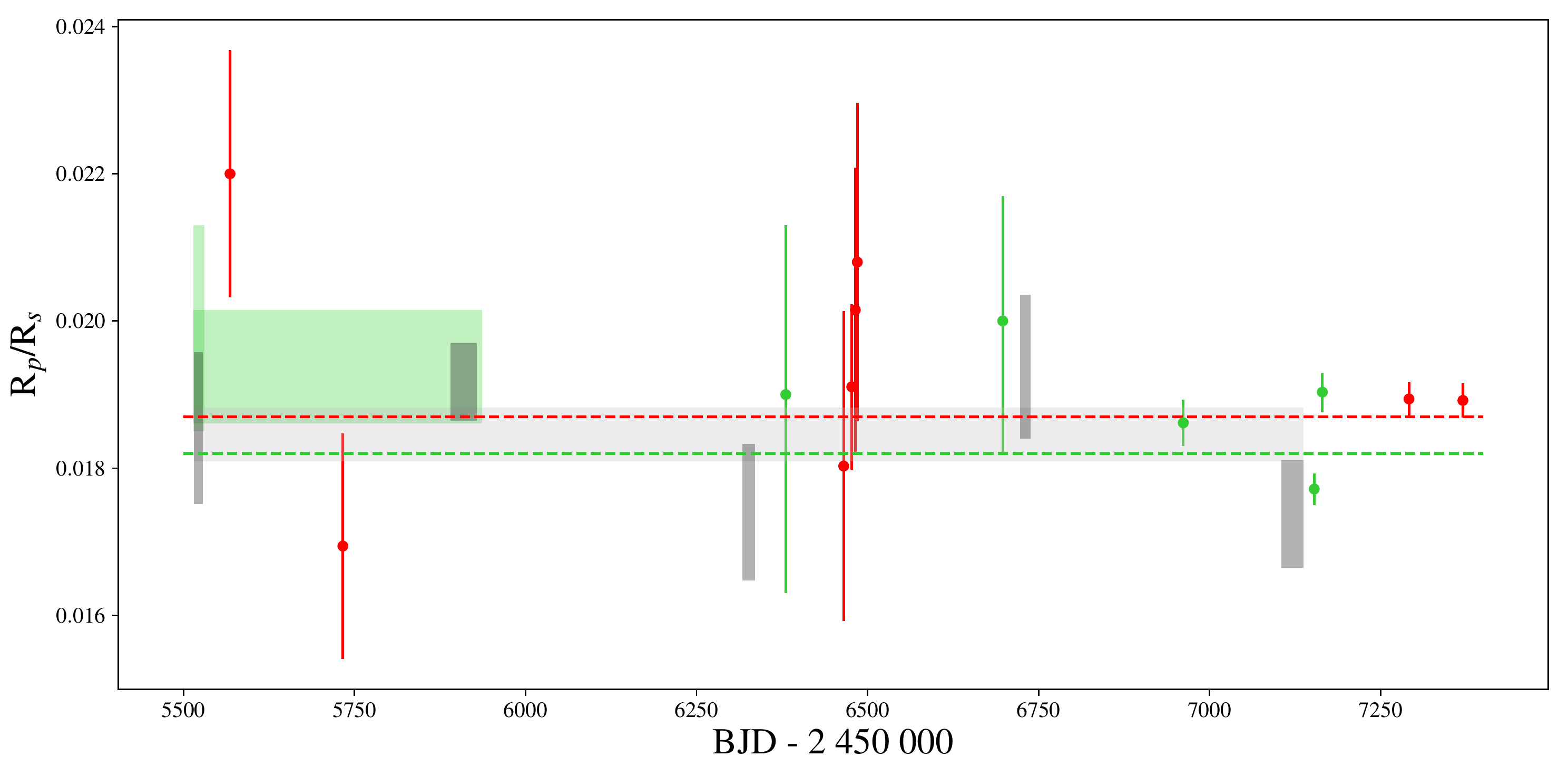}}
    \caption{Reproduction of Fig. 15 of \citetads{2018A&A...619A...1B} representing the planet-to-star radius ratio measured over time with various instruments. 
    Observations in the optical are shown in green. Chronologically, we show values from \citetads{Gillon2} and D14 with MOST; from \citetads{de_Mooij_2014} with Alhambra Faint Object Spectrograph and Camera (ALFOSC) and from \citetads{2018A&A...619A...1B} with the Space Telescope Imaging Spectrograph (STIS) spectrograph onboard the Hubble Space Telescope (HST).
    Our measurements derived from MOST and resulting from the time-shift procedure are shown in black (individual points and combined light curves). 
    Observations in IR are represented in red: \citetads{Demory_2016_nature} with Spitzer and \citetads{Tsiaras_2016} with Wide Field Camera 3 (WFC3) onboard HST.
    Following \citetads{2018A&A...619A...1B}, the dashed green line shows the value obtained from the fit to the three combined STIS visits and the dashed red line shows the value obtained from the fit to the combined Spitzer visits \citepads{Demory_2016}. 
     The values represented by rectangles indicate radius ratios that have been estimated over an extended period of time.}
    \label{Fig15_Vincent}
\end{figure}


\subsection{Search for temporal variations in the transit parameters of 55 Cnc e}

By independently analyzing the five light curves taken from 2011 to 2015, we observe some variations of the transit parameters from year to year. 
This is shown in Fig.~\ref{Fig8_params} for the transit and modulation parameters.
We observe a variability in the transit depth that seems in agreement with \citetads{Demory_2016} and \citetads{2018A&A...619A...1B}, who suggest variations over timescales of days or weeks. This is shown in  Fig.~ \ref{Fig15_Vincent} (reproduction of Fig. 15 of \citeads{2018A&A...619A...1B} with our measurements added). 
However, while we measure a maximum deviation of the transit depth up to $40$ ppm, all the measurements are within the $1~\sigma$ values of each other (see Fig.~\ref{Fig8_params}). Hence, the transit depth variability is not beyond the expected statistical fluctuations of our MOST observations. 

In Fig.~\ref{Fig8_params}, we also compare values obtained before and after the stray-light correction procedures (see  Sec.~\ref{sec32}). As expected, we observe that the stray-light correction improves the precision and even the accuracy of the transit parameters (see top panels), and by extension, the constraint on the eclipse depth. However, the impact on the inferred parameters of the modulation is less pronounced (see bottom panels).

\begin{figure*}[ht] \centering
     \resizebox{0.8\hsize}{!}{\includegraphics{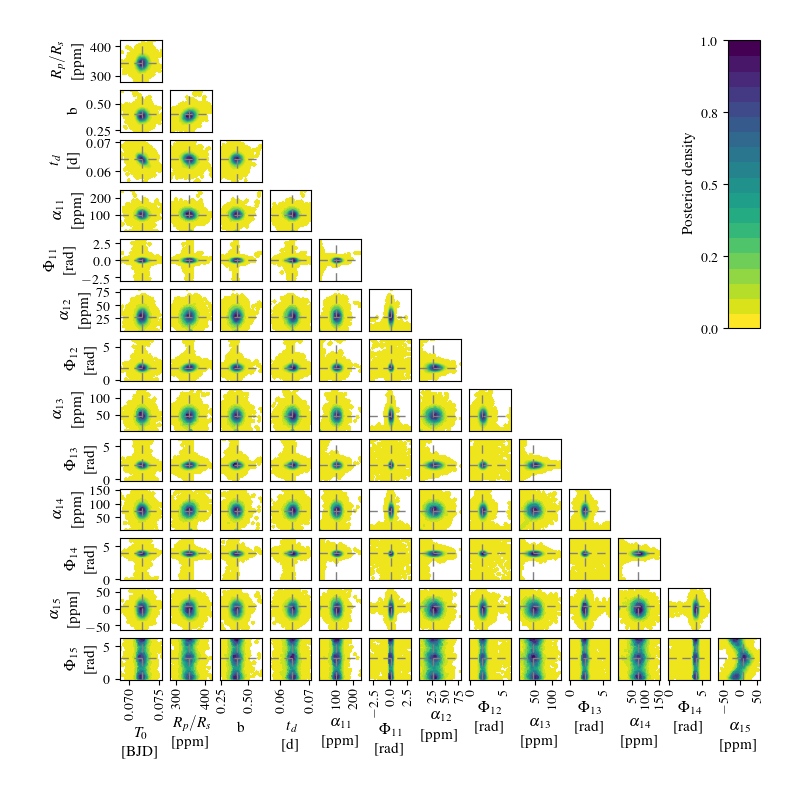}}
    \caption{Pairwise marginal posteriors of the MCMC analysis for the 55 Cnc e  planet and the phase variation modeled as described in \eqref{eq_ec}. For  visibility, we plot $T0-4417$ and fold the $\Phi_{11}$ parameter over $\pm \pi$. We found a double maxima for the $\Phi_{15}$ parameter around $0$ [$\rm mod~2\pi$] and $\pi$, which is meaningless as the corresponding modulation amplitude  $\alpha_{15}$ is roughly null.}
\label{Fig_posterior}
\end{figure*}

\section{Conclusions}
\label{sec:conc}

We analyzed five sequences of MOST observations spanning several weeks of observations taken between 2011 and 2015. We carried out a careful reduction of the raw light curves following current MOST reduction techniques. In addition, we developed a new method, based on the cross-correlation of shorter subsequences to improve the correction of the stray-light noise. 

We performed MCMC analyses of the combined detrended MOST light curves and derived the transit, secondary-eclipse, and phase-modulation parameters. We find transit parameters that are consistent with the most recent values given in \citetads{2018A&A...619A...1B}. The secondary eclipse remains undetected in the MOST observations, but it allows us to constrain the albedo of 55 Cnc~e to an upper limit of $0.47$. We confirm the detection of the optical modulation in flux discovered by \citetads{2011ApJ...737L..18W} at the planet period and detect it at four additional epochs. Intriguingly, we find that its amplitude and phase are variable. At this point, we can only speculate about the origin of the effect. Simultaneous observations at various wavelengths could help to distinguish the origin of this variability that may be linked to the observed modulation observed in far-ultraviolet \citepads{2018A&A...615A.117B} and IR wavelengths \citepads{Demory_2016,Tam18}. We argue that additional observations in the optical with Transiting Exoplanet Survey Satellite (TESS, \citeads{2014SPIE.9143E..20R}) and CHEOPS will be extremely valuable for our understanding of this mysterious planet. Finally, we do not detect any transit for planets b, c, f, and d in the MOST observations.


\begin{acknowledgements}
S. Sulis, M. Lendl, P. Cubillos, and L. Fossati acknowledge support from the Austrian Research Promotion Agency (FFG) under project 859724 ``GRAPPA''. D. Dragomir acknowledges support provided by NASA through Hubble Fellowship grant HSTHF2-51372.001-A awarded by the Space Telescope Science Institute, which is operated by the Association of Universities for Research in Astronomy, Inc., for NASA, under contract NAS5-26555. V. Bourrier acknowledges support by the Swiss National Science Foundation (SNSF) in the frame of the National Centre for Competence in Research PlanetS, and has received funding from the European Research Council (ERC) under the European Union’s Horizon 2020 research and innovation programme (project Four Aces; grant agreement No 724427).
The authors thank Apurva Oza for a careful reading of an earlier version of the manuscript and the anonymous referee for his/her very useful comments.
\end{acknowledgements}

\newpage
\bibliographystyle{aa} 
\bibliography{Bibfile} 

\newcommand{\noop}[1]{}
\begin{thebibliography}{85}
\expandafter\ifx\csname natexlab\endcsname\relax\def\natexlab#1{#1}\fi

\bibitem[{{Adams} {et~al.}(2008){Adams}, {Seager}, \&
  {Elkins-Tanton}}]{2008ApJ...673.1160A}
{Adams}, E.~R., {Seager}, S., \& {Elkins-Tanton}, L. 2008, \apj, 673, 1160

\bibitem[{{Angelo} \& {Hu}(2017)}]{Ang17}
{Angelo}, I. \& {Hu}, R. 2017, \aj, 154, 232

\bibitem[{{Angerhausen} {et~al.}(2015){Angerhausen}, {DeLarme}, \&
  {Morse}}]{Angerhausen2015}
{Angerhausen}, D., {DeLarme}, E., \& {Morse}, J.~A. 2015, \pasp, 127, 1113

\bibitem[{{Batalha} {et~al.}(2011){Batalha}, {Borucki}, {Bryson}, {Buchhave},
  {Caldwell}, {Christensen-Dalsgaard}, {Ciardi}, {Dunham}, {Fressin},
  {Gautier}, {Gilliland}, {Haas}, {Howell}, {Jenkins}, {Kjeldsen}, {Koch},
  {Latham}, {Lissauer}, {Marcy}, {Rowe}, {Sasselov}, {Seager}, {Steffen},
  {Torres}, {Basri}, {Brown}, {Charbonneau}, {Christiansen}, {Clarke},
  {Cochran}, {Dupree}, {Fabrycky}, {Fischer}, {Ford}, {Fortney}, {Girouard},
  {Holman}, {Johnson}, {Isaacson}, {Klaus}, {Machalek}, {Moorehead},
  {Morehead}, {Ragozzine}, {Tenenbaum}, {Twicken}, {Quinn}, {VanCleve},
  {Walkowicz}, {Welsh}, {Devore}, \& {Gould}}]{Bat11}
{Batalha}, N.~M., {Borucki}, W.~J., {Bryson}, S.~T., {et~al.} 2011, \apj, 729,
  27

\bibitem[{{Batygin} \& {Laughlin}(2015)}]{Batygin2015}
{Batygin}, K. \& {Laughlin}, G. 2015, Proceedings of the National Academy of
  Science, 112, 4214

\bibitem[{{Bourrier} {et~al.}(2018{\natexlab{a}}){Bourrier}, {Dumusque},
  {Dorn}, {Henry}, {Astudillo-Defru}, {Rey}, {Benneke}, {H{\'e}brard}, {Lovis},
  {Demory}, {Moutou}, \& {Ehrenreich}}]{2018A&A...619A...1B}
{Bourrier}, V., {Dumusque}, X., {Dorn}, C., {et~al.} 2018{\natexlab{a}}, \aap,
  619, A1

\bibitem[{{Bourrier} {et~al.}(2018{\natexlab{b}}){Bourrier}, {Ehrenreich},
  {Lecavelier des Etangs}, {Louden}, {Wheatley}, {Wyttenbach}, {Vidal-Madjar},
  {Lavie}, {Pepe}, \& {Udry}}]{2018A&A...615A.117B}
{Bourrier}, V., {Ehrenreich}, D., {Lecavelier des Etangs}, A., {et~al.}
  2018{\natexlab{b}}, \aap, 615, A117

\bibitem[{{Broeg} {et~al.}(2013){Broeg}, {Fortier}, {Ehrenreich}, {Alibert},
  {Baumjohann}, {Benz}, {Deleuil}, {Gillon}, {Ivanov}, {Liseau}, {Meyer},
  {Oloffson}, {Pagano}, {Piotto}, {Pollacco}, {Queloz}, {Ragazzoni}, {Renotte},
  {Steller}, \& {Thomas}}]{2013EPJWC..4703005B}
{Broeg}, C., {Fortier}, A., {Ehrenreich}, D., {et~al.} 2013, in European
  Physical Journal Web of Conferences, Vol.~47, European Physical Journal Web
  of Conferences, 03005

\bibitem[{{Cauley} {et~al.}(2018){Cauley}, {Shkolnik}, {Llama}, {Bourrier}, \&
  {Moutou}}]{Cauley_2018}
{Cauley}, P.~W., {Shkolnik}, E.~L., {Llama}, J., {Bourrier}, V., \& {Moutou},
  C. 2018, \aj, 156, 262

\bibitem[{{Chiang} \& {Laughlin}(2013)}]{Chiang2013}
{Chiang}, E. \& {Laughlin}, G. 2013, \mnras, 431, 3444

\bibitem[{{Crossfield}(2012)}]{Cro12}
{Crossfield}, I.~J.~M. 2012, \aap, 545, A97

\bibitem[{{Cubillos} {et~al.}(2017){Cubillos}, {Harrington}, {Loredo}, {Lust},
  {Blecic}, \& {Stemm}}]{2017AJ....153....3C}
{Cubillos}, P., {Harrington}, J., {Loredo}, T.~J., {et~al.} 2017, \aj, 153, 3

\bibitem[{{Davis} \& {Wheatley}(2009)}]{Davis_2009}
{Davis}, T.~A. \& {Wheatley}, P.~J. 2009, \mnras, 396, 1012

\bibitem[{{Dawson} \& {Fabrycky}(2010)}]{Dawson}
{Dawson}, R.~I. \& {Fabrycky}, D.~C. 2010, ApJ, 722, 937

\bibitem[{de~Mooij {et~al.}(2014)de~Mooij, L{\'{o}}pez-Morales, Karjalainen,
  Hrudkova, \& Jayawardhana}]{de_Mooij_2014}
de~Mooij, E. J.~W., L{\'{o}}pez-Morales, M., Karjalainen, R., Hrudkova, M., \&
  Jayawardhana, R. 2014, ApJ, 797, L21

\bibitem[{{Demory}(2014)}]{Dem14}
{Demory}, B.-O. 2014, \apjl, 789, L20

\bibitem[{{Demory} {et~al.}(2016{\natexlab{a}}){Demory}, {Gillon}, {de Wit},
  {Madhusudhan}, {Bolmont}, {Heng}, {Kataria}, {Lewis}, {Hu}, {Krick},
  {Stamenkovi{\'c}}, {Benneke}, {Kane}, \& {Queloz}}]{Demory_2016_nature}
{Demory}, B.-O., {Gillon}, M., {de Wit}, J., {et~al.} 2016{\natexlab{a}}, \nat,
  532, 207

\bibitem[{{Demory} {et~al.}(2011{\natexlab{a}}){Demory}, {Gillon}, {Deming},
  {Valencia}, {Seager}, {Benneke}, {Lovis}, {Cubillos}, {Harrington},
  {Stevenson}, {Mayor}, {Pepe}, {Queloz}, {S{\'e}gransan}, \& {Udry}}]{Dem11}
{Demory}, B.-O., {Gillon}, M., {Deming}, D., {et~al.} 2011{\natexlab{a}}, A\&A,
  533, A114

\bibitem[{{Demory} {et~al.}(2016{\natexlab{b}}){Demory}, {Gillon},
  {Madhusudhan}, \& {Queloz}}]{Demory_2016}
{Demory}, B.-O., {Gillon}, M., {Madhusudhan}, N., \& {Queloz}, D.
  2016{\natexlab{b}}, \mnras, 455, 2018

\bibitem[{{Demory} {et~al.}(2011{\natexlab{b}}){Demory}, {Seager},
  {Madhusudhan}, {Kjeldsen}, {Christensen-Dalsgaard}, {Gillon}, {Rowe},
  {Welsh}, {Adams}, {Dupree}, {McCarthy}, {Kulesa}, {Borucki}, \&
  {Koch}}]{Demory11b}
{Demory}, B.-O., {Seager}, S., {Madhusudhan}, N., {et~al.} 2011{\natexlab{b}},
  \apjl, 735, L12

\bibitem[{{Di Mauro}(2016)}]{DiMauro2016}
{Di Mauro}, M.~P. 2016, in Frontier Research in Astrophysics II, 29

\bibitem[{Dorn {et~al.}(2018)Dorn, Harrison, Bonsor, \& Hands}]{Dorn18}
Dorn, C., Harrison, J. H.~D., Bonsor, A., \& Hands, T.~O. 2018, MNRAS, 484, 712

\bibitem[{{Dragomir} {et~al.}(2013){Dragomir}, {Matthews}, {Eastman},
  {Cameron}, {Howard}, {Guenther}, {Kuschnig}, {Moffat}, {Rowe}, {Rucinski},
  {Sasselov}, \& {Weiss}}]{Dra13}
{Dragomir}, D., {Matthews}, J.~M., {Eastman}, J.~D., {et~al.} 2013, \apjl, 772,
  L2

\bibitem[{{Dragomir} {et~al.}(2014){Dragomir}, {Matthews}, {Winn}, \&
  {Rowe}}]{2014IAUS..293...52D}
{Dragomir}, D., {Matthews}, J.~M., {Winn}, J.~N., \& {Rowe}, J.~F. 2014, in IAU
  Symposium, Vol. 293, IAU Symposium, ed. N.~{Haghighipour}, 52--57

\bibitem[{{Ehrenreich} {et~al.}(2012){Ehrenreich}, {Bourrier}, {Bonfils},
  {Lecavelier des Etangs}, {H{\'e}brard}, {Sing}, {Wheatley}, {Vidal-Madjar},
  {Delfosse}, {Udry}, {Forveille}, \& {Moutou}}]{Ehr12}
{Ehrenreich}, D., {Bourrier}, V., {Bonfils}, X., {et~al.} 2012, \aap, 547, A18

\bibitem[{{Ehrenreich, D.} \& {D\'esert, J.-M.}(2011)}]{Ehrenreich_2011}
{Ehrenreich, D.} \& {D\'esert, J.-M.} 2011, A\&A, 529, A136

\bibitem[{{Esteves} {et~al.}(2015){Esteves}, {De Mooij}, \&
  {Jayawardhana}}]{Esteves2015}
{Esteves}, L.~J., {De Mooij}, E. J.~W., \& {Jayawardhana}, R. 2015, \apj, 804,
  150

\bibitem[{{Fischer} {et~al.}(2008){Fischer}, {Marcy}, {Butler}, {Vogt},
  {Laughlin}, {Henry}, {Abouav}, {Peek}, {Wright}, {Johnson}, {McCarthy}, \&
  {Isaacson}}]{Fischer1}
{Fischer}, D.~A., {Marcy}, G.~W., {Butler}, R.~P., {et~al.} 2008, ApJ, 675, 790

\bibitem[{{Folsom} {et~al.}(\noop{2019}submitted){Folsom}, {Fionnagáin},
  {Fossati}, {Vidotto}, {Moutou}, {Petit}, {Dragomir}, \&
  {Donati}}]{Folsom-inpress-a}
{Folsom}, C.~P., {Fionnagáin}, D.~O., {Fossati}, L., {et~al.}
  \noop{2019}submitted

\bibitem[{{Fressin} {et~al.}(2013){Fressin}, {Torres}, {Charbonneau}, {Bryson},
  {Christiansen}, {Dressing}, {Jenkins}, {Walkowicz}, \& {Batalha}}]{Fressin13}
{Fressin}, F., {Torres}, G., {Charbonneau}, D., {et~al.} 2013, \apj, 766, 81

\bibitem[{{Fulton} \& {Petigura}(2018)}]{Fulton_2018}
{Fulton}, B.~J. \& {Petigura}, E.~A. 2018, \aj, 156, 264

\bibitem[{{Fulton} {et~al.}(2017){Fulton}, {Petigura}, {Howard}, {Isaacson},
  {Marcy}, {Cargile}, {Hebb}, {Weiss}, {Johnson}, {Morton}, {Sinukoff},
  {Crossfield}, \& {Hirsch}}]{Fulton17}
{Fulton}, B.~J., {Petigura}, E.~A., {Howard}, A.~W., {et~al.} 2017, \aj, 154,
  109

\bibitem[{{Gandolfi} {et~al.}(2018){Gandolfi}, {Barrag{\'a}n}, {Livingston},
  {Fridlund}, {Justesen}, {Redfield}, {Fossati}, {Mathur}, {Grziwa}, {Cabrera},
  {Garc{\'\i}a}, {Persson}, {Van Eylen}, {Hatzes}, {Hidalgo}, {Albrecht},
  {Bugnet}, {Cochran}, {Csizmadia}, {Deeg}, {Eigm{\"u}ller}, {Endl}, {Erikson},
  {Esposito}, {Guenther}, {Korth}, {Luque}, {Monta{\~n}es Rodr{\'\i}guez},
  {Nespral}, {Nowak}, {P{\"a}tzold}, \& {Prieto-Arranz}}]{Gandolfi_2018}
{Gandolfi}, D., {Barrag{\'a}n}, O., {Livingston}, J.~H., {et~al.} 2018, \aap,
  619, L10

\bibitem[{{Gillon} {et~al.}(2012){Gillon}, {Demory}, {Benneke}, {Valencia},
  {Deming}, {Seager}, {Lovis}, {Mayor}, {Pepe}, {Queloz}, {S{\'e}gransan}, \&
  {Udry}}]{Gillon2}
{Gillon}, M., {Demory}, B.-O., {Benneke}, B., {et~al.} 2012, A\&A, 539, A28

\bibitem[{{Gillon} {et~al.}(2010){Gillon}, {Lanotte}, {Barman}, {Miller},
  {Demory}, {Deleuil}, {Montalb{\'a}n}, {Bouchy}, {Collier Cameron}, {Deeg},
  {Fortney}, {Fridlund}, {Harrington}, {Magain}, {Moutou}, {Queloz}, {Rauer},
  {Rouan}, \& {Schneider}}]{2010A&A...511A...3G}
{Gillon}, M., {Lanotte}, A.~A., {Barman}, T., {et~al.} 2010, \aap, 511, A3

\bibitem[{{Hansen} \& {Murray}(2012)}]{Hansen2012}
{Hansen}, B. M.~S. \& {Murray}, N. 2012, \apj, 751, 158

\bibitem[{{Jansen} \& {Kipping}(2018)}]{Jan18}
{Jansen}, T. \& {Kipping}, D. 2018, \mnras, 478, 3025

\bibitem[{{Jin} \& {Mordasini}(2018)}]{jin2018}
{Jin}, S. \& {Mordasini}, C. 2018, \apj, 853, 163

\bibitem[{{Jones} {et~al.}(2002){Jones}, {Paul Butler}, {Tinney}, {Marcy},
  {Penny}, {McCarthy}, {Carter}, \& {Pourbaix}}]{Hugh_2002}
{Jones}, H. R.~A., {Paul Butler}, R., {Tinney}, C.~G., {et~al.} 2002, MNRAS,
  333, 871

\bibitem[{{Kane} {et~al.}(2011){Kane}, {Gelino}, {Ciardi}, {Dragomir}, \& {von
  Braun}}]{2011ApJ...740...61K}
{Kane}, S.~R., {Gelino}, D.~M., {Ciardi}, D.~R., {Dragomir}, D., \& {von
  Braun}, K. 2011, \apj, 740, 61

\bibitem[{{Kane} {et~al.}(2016){Kane}, {Wittenmyer}, {Hinkel}, {Roy},
  {Mahadevan}, {Dragomir}, {Matthews}, {Henry}, {Chakraborty}, {Boyajian},
  {Wright}, {Ciardi}, {Fischer}, {Butler}, {Tinney}, {Carter}, {Jones},
  {Bailey}, \& {O'Toole}}]{2016ApJ...821...65K}
{Kane}, S.~R., {Wittenmyer}, R.~A., {Hinkel}, N.~R., {et~al.} 2016, \apj, 821,
  65

\bibitem[{{Kite} {et~al.}(2016){Kite}, {Fegley}, {Schaefer}, \&
  {Gaidos}}]{Kite16}
{Kite}, E.~S., {Fegley}, Jr., B., {Schaefer}, L., \& {Gaidos}, E. 2016, \apj,
  828, 80

\bibitem[{{Kr{\"u}ger} {et~al.}(2003){Kr{\"u}ger}, {Geissler}, {Hor{\'a}nyi},
  {Graps}, {Kempf}, {Srama}, {Moragas-Klostermeyer}, {Moissl}, {Johnson}, \&
  {Gr{\"u}n}}]{2003GeoRL..30.2101K}
{Kr{\"u}ger}, H., {Geissler}, P., {Hor{\'a}nyi}, M., {et~al.} 2003, \grl, 30,
  2101

\bibitem[{{Kubyshkina} {et~al.}(2018){Kubyshkina}, {Fossati}, {Erkaev},
  {Johnstone}, {Cubillos}, {Kislyakova}, {Lammer}, {Lendl}, \&
  {Odert}}]{Kubyshkina18}
{Kubyshkina}, D., {Fossati}, L., {Erkaev}, N.~V., {et~al.} 2018, \aap, 619,
  A151

\bibitem[{{Lanza}(2012)}]{Lanza_2012}
{Lanza}, A.~F. 2012, \aap, 544, A23

\bibitem[{{Lecavelier des Etangs, A.}(2007)}]{Lecavelier_2007}
{Lecavelier des Etangs, A.} 2007, A\&A, 461, 1185

\bibitem[{{Lendl} {et~al.}(2017){Lendl}, {Cubillos}, {Hagelberg}, {M{\"u}ller},
  {Juvan}, \& {Fossati}}]{2017A&A...606A..18L}
{Lendl}, M., {Cubillos}, P.~E., {Hagelberg}, J., {et~al.} 2017, \aap, 606, A18

\bibitem[{{Maggio} {et~al.}(2015){Maggio}, {Pillitteri}, {Scandariato},
  {Lanza}, {Sciortino}, {Borsa}, {Bonomo}, {Claudi}, {Covino}, {Desidera},
  {Gratton}, {Micela}, {Pagano}, {Piotto}, {Sozzetti}, {Cosentino}, \&
  {Maldonado}}]{Maggio_2015}
{Maggio}, A., {Pillitteri}, I., {Scandariato}, G., {et~al.} 2015, \apjl, 811,
  L2

\bibitem[{{Mandel} \& {Agol}(2002)}]{2002ApJ...580L.171M}
{Mandel}, K. \& {Agol}, E. 2002, \apjl, 580, L171

\bibitem[{{Matthews}(2004)}]{2004AAS...20513401M}
{Matthews}, J.~M. 2004, in BAAS, Vol.~36, AAS Meeting Abstracts, 1563

\bibitem[{{Mayor} {et~al.}(2011){Mayor}, {Marmier}, {Lovis}, {Udry},
  {S{\'e}gransan}, {Pepe}, {Benz}, {Bertaux}, {Bouchy}, {Dumusque}, {Lo Curto},
  {Mordasini}, {Queloz}, \& {Santos}}]{Mayor11}
{Mayor}, M., {Marmier}, M., {Lovis}, C., {et~al.} 2011, arXiv e-prints,
  arXiv:1109.2497

\bibitem[{{Miguel}(2019)}]{Mig19}
{Miguel}, Y. 2019, \mnras, 482, 2893

\bibitem[{{Mordasini} {et~al.}(2012){Mordasini}, {Alibert}, {Georgy},
  {Dittkrist}, {Klahr}, \& {Henning}}]{Mordasini12}
{Mordasini}, C., {Alibert}, Y., {Georgy}, C., {et~al.} 2012, A\&A, 547, A112

\bibitem[{{Motalebi} {et~al.}(2015){Motalebi}, {Udry}, {Gillon}, {Lovis},
  {S{\'e}gransan}, {Buchhave}, {Demory}, {Malavolta}, {Dressing}, {Sasselov},
  {Rice}, {Charbonneau}, {Collier Cameron}, {Latham}, {Molinari}, {Pepe},
  {Affer}, {Bonomo}, {Cosentino}, {Dumusque}, {Figueira}, {Fiorenzano},
  {Gettel}, {Harutyunyan}, {Haywood}, {Johnson}, {Lopez}, {Lopez-Morales},
  {Mayor}, {Micela}, {Mortier}, {Nascimbeni}, {Philips}, {Piotto}, {Pollacco},
  {Queloz}, {Sozzetti}, {Vanderburg}, \& {Watson}}]{Motalebi_2015}
{Motalebi}, F., {Udry}, S., {Gillon}, M., {et~al.} 2015, \aap, 584, A72

\bibitem[{{Namekata} {et~al.}(2019){Namekata}, {Maehara}, {Notsu}, {Toriumi},
  {Hayakawa}, {Ikuta}, {Notsu}, {Honda}, {Nogami}, \& {Shibata}}]{Namekata19}
{Namekata}, K., {Maehara}, H., {Notsu}, Y., {et~al.} 2019, \apj, 871, 187

\bibitem[{{Owen} \& {Wu}(2017)}]{owen2017}
{Owen}, J.~E. \& {Wu}, Y. 2017, \apj, 847, 29

\bibitem[{{Pagano} {et~al.}(2009){Pagano}, {Lanza}, {Leto}, {Messina}, {Barge},
  \& {Baglin}}]{Pagano_2009}
{Pagano}, I., {Lanza}, A.~F., {Leto}, G., {et~al.} 2009, Earth Moon and
  Planets, 105, 373

\bibitem[{{Petigura} {et~al.}(2013){Petigura}, {Howard}, \&
  {Marcy}}]{Petigura13}
{Petigura}, E.~A., {Howard}, A.~W., \& {Marcy}, G.~W. 2013, Proceedings of the
  National Academy of Science, 110, 19273

\bibitem[{{Pont} {et~al.}(2006){Pont}, {Zucker}, \&
  {Queloz}}]{2006MNRAS.373..231P}
{Pont}, F., {Zucker}, S., \& {Queloz}, D. 2006, \mnras, 373, 231

\bibitem[{{Poppenhaeger} \& {Schmitt}(2011)}]{Poppenhaeger_2011}
{Poppenhaeger}, K. \& {Schmitt}, J.~H.~M.~M. 2011, \apj, 735, 59

\bibitem[{{Ricker} {et~al.}(2014){Ricker}, {Winn}, {Vanderspek}, {Latham},
  {Bakos}, {Bean}, {Berta-Thompson}, {Brown}, {Buchhave}, {Butler}, {Butler},
  {Chaplin}, {Charbonneau}, {Christensen-Dalsgaard}, {Clampin}, {Deming},
  {Doty}, {De Lee}, {Dressing}, {Dunham}, {Endl}, {Fressin}, {Ge}, {Henning},
  {Holman}, {Howard}, {Ida}, {Jenkins}, {Jernigan}, {Johnson}, {Kaltenegger},
  {Kawai}, {Kjeldsen}, {Laughlin}, {Levine}, {Lin}, {Lissauer}, {MacQueen},
  {Marcy}, {McCullough}, {Morton}, {Narita}, {Paegert}, {Palle}, {Pepe},
  {Pepper}, {Quirrenbach}, {Rinehart}, {Sasselov}, {Sato}, {Seager},
  {Sozzetti}, {Stassun}, {Sullivan}, {Szentgyorgyi}, {Torres}, {Udry}, \&
  {Villasenor}}]{2014SPIE.9143E..20R}
{Ricker}, G.~R., {Winn}, J.~N., {Vanderspek}, R., {et~al.} 2014, in Society of
  Photo-Optical Instrumentation Engineers (SPIE) Conference Series, Vol. 9143,
  \procspie, 914320

\bibitem[{{Ridden-Harper} {et~al.}(2016){Ridden-Harper}, {Snellen}, {Keller},
  {de Kok}, {Di Gloria}, {Hoeijmakers}, {Brogi}, {Fridlund}, {Vermeersen}, \&
  {van Westrenen}}]{Rid16}
{Ridden-Harper}, A.~R., {Snellen}, I.~A.~G., {Keller}, C.~U., {et~al.} 2016,
  \aap, 593, A129

\bibitem[{{Rogers} {et~al.}(2011){Rogers}, {Bodenheimer}, {Lissauer}, \&
  {Seager}}]{Rogers2011}
{Rogers}, L.~A., {Bodenheimer}, P., {Lissauer}, J.~J., \& {Seager}, S. 2011,
  \apj, 738, 59

\bibitem[{{Rogers} \& {Seager}(2010)}]{Rogers1}
{Rogers}, L.~A. \& {Seager}, S. 2010, ApJ, 712, 974

\bibitem[{{Rouan} {et~al.}(2011){Rouan}, {Deeg}, {Demangeon}, {Samuel},
  {Cavarroc}, {Fegley}, \& {L{\'e}ger}}]{Rou11}
{Rouan}, D., {Deeg}, H.~J., {Demangeon}, O., {et~al.} 2011, \apjl, 741, L30

\bibitem[{{Rowe} {et~al.}(2006){Rowe}, {Matthews}, {Seager}, {Kuschnig},
  {Guenther}, {Moffat}, {Rucinski}, {Sasselov}, {Walker}, \&
  {Weiss}}]{2006ApJ...646.1241R}
{Rowe}, J.~F., {Matthews}, J.~M., {Seager}, S., {et~al.} 2006, \apj, 646, 1241

\bibitem[{{Rowe} {et~al.}(2008){Rowe}, {Matthews}, {Seager}, {Miller-Ricci},
  {Sasselov}, {Kuschnig}, {Guenther}, {Moffat}, {Rucinski}, {Walker}, \&
  {Weiss}}]{Row08}
{Rowe}, J.~F., {Matthews}, J.~M., {Seager}, S., {et~al.} 2008, \apj, 689, 1345

\bibitem[{{Rucinski} {et~al.}(2004)}]{2004PASP..116.1093R}
{Rucinski}, S.~M. {et~al.} 2004, \pasp, 116, 1093

\bibitem[{Savitzky \& Golay(1964)}]{savitzky64}
Savitzky, A. \& Golay, M. J.~E. 1964, Analytical Chemistry, 36, 1627

\bibitem[{{Sheets} \& {Deming}(2017)}]{She17}
{Sheets}, H.~A. \& {Deming}, D. 2017, \aj, 154, 160

\bibitem[{{Shkolnik} {et~al.}(2003){Shkolnik}, {Walker}, \&
  {Bohlender}}]{Shkolnik_2003}
{Shkolnik}, E., {Walker}, G.~A.~H., \& {Bohlender}, D.~A. 2003, \apj, 597, 1092

\bibitem[{{Shkolnik} \& {Llama}(2018)}]{Shkolnik_2018}
{Shkolnik}, E.~L. \& {Llama}, J. 2018, {Signatures of Star-Planet
  Interactions}, 20

\bibitem[{{Strugarek} {et~al.}(2019){Strugarek}, {Brun}, {Donati}, {Moutou}, \&
  {R{\'e}ville}}]{Strugarek_2019}
{Strugarek}, A., {Brun}, A.~S., {Donati}, J.~F., {Moutou}, C., \&
  {R{\'e}ville}, V. 2019, arXiv e-prints, arXiv:1907.01020

\bibitem[{{Strugarek} {et~al.}(2015){Strugarek}, {Brun}, {Matt}, \&
  {R{\'e}ville}}]{Strugarek_2015}
{Strugarek}, A., {Brun}, A.~S., {Matt}, S.~P., \& {R{\'e}ville}, V. 2015, \apj,
  815, 111

\bibitem[{{Tamburo} {et~al.}(2018){Tamburo}, {Mandell}, {Deming}, \&
  {Garhart}}]{Tam18}
{Tamburo}, P., {Mandell}, A., {Deming}, D., \& {Garhart}, E. 2018, \aj, 155,
  221

\bibitem[{Tsiaras {et~al.}(2016)Tsiaras, Rocchetto, Waldmann, Venot, Varley,
  Morello, Damiano, Tinetti, Barton, Yurchenko, \& Tennyson}]{Tsiaras_2016}
Tsiaras, A., Rocchetto, M., Waldmann, I.~P., {et~al.} 2016, AJ, 820, 99

\bibitem[{{Van Eylen} {et~al.}(2018){Van Eylen}, {Agentoft}, {Lundkvist},
  {Kjeldsen}, {Owen}, {Fulton}, {Petigura}, \& {Snellen}}]{Eyl17}
{Van Eylen}, V., {Agentoft}, C., {Lundkvist}, M.~S., {et~al.} 2018, \mnras,
  479, 4786

\bibitem[{{Vogt} {et~al.}(2015){Vogt}, {Burt}, {Meschiari}, {Butler}, {Henry},
  {Wang}, {Holden}, {Gapp}, {Hanson}, {Arriagada}, {Keiser}, {Teske}, \&
  {Laughlin}}]{Vogt_2015}
{Vogt}, S.~S., {Burt}, J., {Meschiari}, S., {et~al.} 2015, \apj, 814, 12

\bibitem[{{von Braun} {et~al.}(2011){von Braun}, {Boyajian}, {ten Brummelaar},
  {Kane}, {van Belle}, {Ciardi}, {Raymond}, {L{\'o}pez-Morales}, {McAlister},
  {Schaefer}, {Ridgway}, {Sturmann}, {Sturmann}, {White}, {Turner},
  {Farrington}, \& {Goldfinger}}]{Braun3}
{von Braun}, K., {Boyajian}, T.~S., {ten Brummelaar}, T.~A., {et~al.} 2011,
  ApJ, 740, 49

\bibitem[{{Walker} {et~al.}(2003){Walker}, {Matthews}, {Kuschnig}, {Johnson},
  {Rucinski}, {Pazder}, {Burley}, {Walker}, {Skaret}, {Zee}, {Grocott},
  {Carroll}, {Sinclair}, {Sturgeon}, \& {Harron}}]{2003PASP..115.1023W}
{Walker}, G., {Matthews}, J., {Kuschnig}, R., {et~al.} 2003, \pasp, 115, 1023

\bibitem[{{Walker} {et~al.}(2008){Walker}, {Croll}, {Matthews}, {Kuschnig},
  {Huber}, {Weiss}, {Shkolnik}, {Rucinski}, {Guenther}, {Moffat}, \&
  {Sasselov}}]{Walker_2008}
{Walker}, G.~A.~H., {Croll}, B., {Matthews}, J.~M., {et~al.} 2008, \aap, 482,
  691

\bibitem[{{Winn} {et~al.}(2008){Winn}, {Holman}, {Torres}, {McCullough},
  {Johns-Krull}, {Latham}, {Shporer}, {Mazeh}, {Garcia-Melendo}, {Foote},
  {Esquerdo}, \& {Everett}}]{2008ApJ...683.1076W}
{Winn}, J.~N., {Holman}, M.~J., {Torres}, G., {et~al.} 2008, \apj, 683, 1076

\bibitem[{{Winn} {et~al.}(2011){Winn}, {Matthews}, {Dawson}, {Fabrycky},
  {Holman}, {Kallinger}, {Kuschnig}, {Sasselov}, {Dragomir}, {Guenther},
  {Moffat}, {Rowe}, {Rucinski}, \& {Weiss}}]{2011ApJ...737L..18W}
{Winn}, J.~N., {Matthews}, J.~M., {Dawson}, R.~I., {et~al.} 2011, \apjl, 737,
  L18

\bibitem[{{Wright} \& {Miller}(2015)}]{Wright_2015}
{Wright}, J.~T. \& {Miller}, B. 2015, in IAU General Assembly, Vol.~29, 2258453

\bibitem[{{Yee} {et~al.}(2017){Yee}, {Petigura}, \& {von
  Braun}}]{2017ApJ...836...77Y}
{Yee}, S.~W., {Petigura}, E.~A., \& {von Braun}, K. 2017, \apj, 836, 77

\end{thebibliography}


\begin{appendix}

 \section{Long-term variability affecting the 55 Cancri light curve: Signature of the stellar activity}
\label{App1}

\begin{figure*}[!b] \centering
     \resizebox{\hsize}{!}{\includegraphics{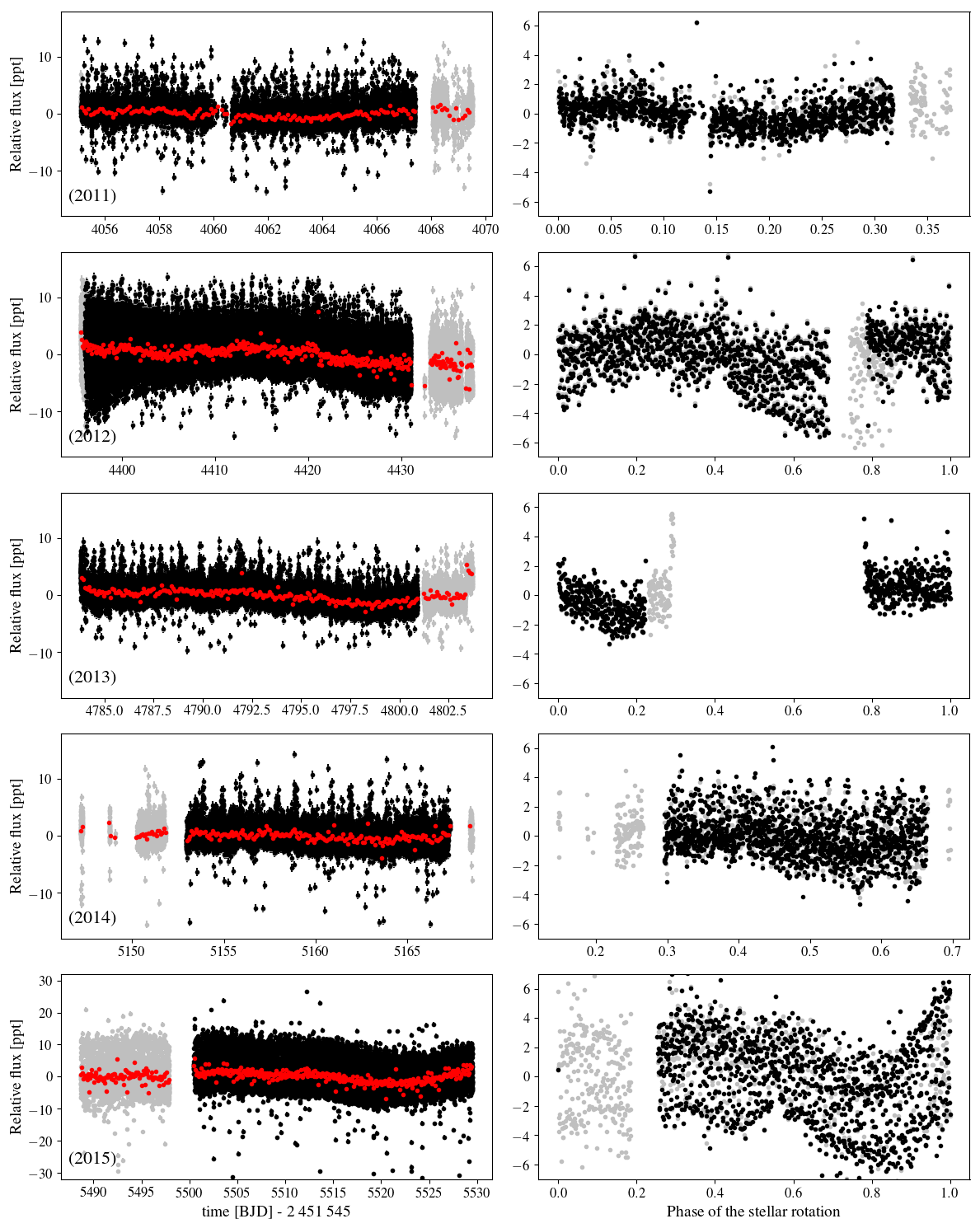}}
    \caption{From top to bottom: Light curves taken in 2011, 2012, 2013, 2014, and 2015 obtained before the long-term variation correction in the pre-whitening stage (see Sec.~\ref{sec31}). The y-axis is in part-per-thousand. The removed sections of the light curves are shown in gray. The left column panels represent the light curve as a function of time. The long-term variation is well observed on the data binned at two-hour intervals (red). The right column panels represent these flux phase folded at the stellar rotation period of $38.8$ days (binned into $30$-min intervals). We note the different scale on the y-axis of the 2015 plot on the bottom. }
\label{Fig_App}
\end{figure*}

During the pre-whitening stage (see Sec.~\ref{sec31}), we observe a long-term variation in each of the light curves. This variation is shown in Fig.~\ref{Fig_App}. We observe variations over timescales close to half of the stellar rotation period ($38.8/2$ day) but that changes in both amplitude and phase from year to year (see right column). When binning the observations into two-hour intervals (see left column), we measure a dispersion around the mean value of $656$, $1471$, $1116$, $774,$ and $1733$ ppm for the 2011 to 2015 datasets, respectively. 
We attribute these variations to the evolution of the stellar activity as they are not strictly increasing as would be expected in case of instrumental systematics. We note that the stray-light systematics discussed in Sec.~\ref{sec32} are clearly visible in the three first datasets. Correcting for the long-term variation before correcting the stray-light systematics, as done in this study contrary to previous studies of W11 and D14, avoids the introduction of an offset between the one-day sequences used in the analysis described in Sec.~\ref{sec32}.

\end{appendix}

\end{document}